\documentclass[convention,peer-reviewed]{aesconf}

\graphicspath{{./}{figures/}}

\usepackage[utf8]{inputenc}

\usepackage{microtype}

\usepackage[numbers,square]{natbib}
\usepackage{url}

\usepackage{booktabs}
\usepackage{color}
\usepackage{url}
\usepackage{graphicx}
\usepackage{subcaption}
\usepackage[margin=1in]{geometry}

\title{Measurements and Analysis of Distortion in Inertial Force Exciters for Flat-Panel Loudspeakers}

\author[1]{Stephen F. Roessner}
\author[1]{Michael C. Heilemann}

\affil[1]{Department of Sound Arts and Engineering, University of Rochester, Rochester, NY 14127}

\correspondence{Stephen F. Roessner}{stephen.roessner@rochester.edu}

\lastnames{Roessner and Heilemann}

\shorttitle{Distortion in Force Exciters}

\begin{document}

\twocolumn[
\maketitle 

\begin{onecolabstract}

Flat-panel loudspeakers are typically driven by dynamic exciters that employ a voice coil coupled to a reaction mass to exert force on the panel. Unlike conventional loudspeakers, where the cone is relatively light weight, heavy panels may require significant exertion from the driver to achieve sound pressure levels required of consumer grade loudspeakers. Overexertion can cause harmonic distortion and reduced sound quality, especially in drivers where the voice coil is not centered in the magnetic field. There are many different exciters available commercially with different power ratings and price points. The frequency response, linearity, total-harmonic distortion (THD), intermodulation distortion (IMD), and force vs. displacement were measured for three types of exciters to determine the consistency of manufacturing and use in panel loudspeakers. These exciters were found to have high variability in all metrics among units of the same type. Distortion metrics such as IMD may be reduced by running impedance sweeps to match exciters with similar response characteristics in multi-actuator panels. 

\end{onecolabstract}
]
\section{Introduction}



 Force exciters made for driving flat-panel loudspeakers (FPLs) are typically designed much like traditional speakers, with a magnet, voice coil, and surround or spider. These elements employ the inertia of a coupled reaction mass to excite the bending waves in the speaker that radiate sound. This has served as the working principle behind many iterations of flat-panel speakers from early prototypes \cite{lane1930sound,Hiromi}, to distributed mode loudspeakers \cite{BankHarris,AzimaPatent}, to more modern designs that may employ multiple actuators \cite{anderson2017flat-panel,anderson2017impulse,Cecchi2019applying,Zenker2019rethinking} as described in \cite{heilemann2021evolution}. Several listening tests have been performed on panel loudspeakers since their inception \cite{HeilemannListenerPreference,Flanagan1999,HarrisThesis}. Though some designs employing multiple exciters \cite{roessner2019evaluating} or equalization \cite{10.1121/10.0036373} have received positive reviews, panel loudspeakers as a whole tend to lag behind conventional speakers in terms of listener preference. 

 One reason for this may be the loudspeaker distortion, which is closely related to the excursion of the diaphragm or panel \cite{klipsch1972modulation}. Distortion measurements in traditional loudspeakers have been shown to have no correlation with listener preference \cite{olive2004multiple1,olive2004multiple2}. This is true because distortion levels are usually below the threshold of audibility in consumer grade loudspeakers, meaning that listeners cannot distinguish a loudspeaker with higher distortion from one with lower distortion. The radiating surface used in FPLs however, are much heavier than the speaker cones in conventional loudspeakers and requires the exciters to be pushed harder to achieve comparable sound pressure levels (SPLs) across the audio bandwidth, thus increasing the potential for harmonic distortion above the threshold of audibility. In addition, exciters are affixed to the panel using an adhesive attached to the voice coil only. The reaction mass is typically unbraced, which can cause rocking motion when excited on resonance. This can cause several other non-linear behaviors to appear. An analysis of exciter responses and distortion characteristics will allow users to select the right exciter for their FPL designs to achieve the highest sound quality. We have adopted typical loudspeaker performance tests to characterize the performance of different commercially available exciters for FPLs \cite{beers1943frequency, metzler1993, klipsch1969modulation,klipsch1970modulation,klipsch1972modulation, klippel2001speaker, klippel2005loudspeaker, klippel2007optimal}.


\section{Experimental Setup}
Like pistonic speakers, there are many models, sizes, and power ratings of contact exciters that can be utilized for FPLs. Three types of exciters were tested in this study: the Dayton Audio DAEX25FHE-4 (Fig.~\ref{fig:DAEX25FHE-4_Back}), the Dayton Audio DAEX25CT-4 (Fig.~\ref{fig:DAEX25CT-4_Back}), and the Billionsound 12K09 BS12K09 (Fig.~\ref{fig:BS12K09_Back}). 

The manufacturer specifications for each are listed in Table~\ref{table:exciterspecs}. Multiple individual exciters were tested for each make and model to determine the characteristic variability. Note that only five BS12K09 exciters were tested due to supply-chain shortages at the time of this study.

\begin{figure}[t]
\captionsetup{font=small}
  \centering
  \begin{subfigure}[b]{0.32\linewidth}
    \centering
    \includegraphics[width=\linewidth]{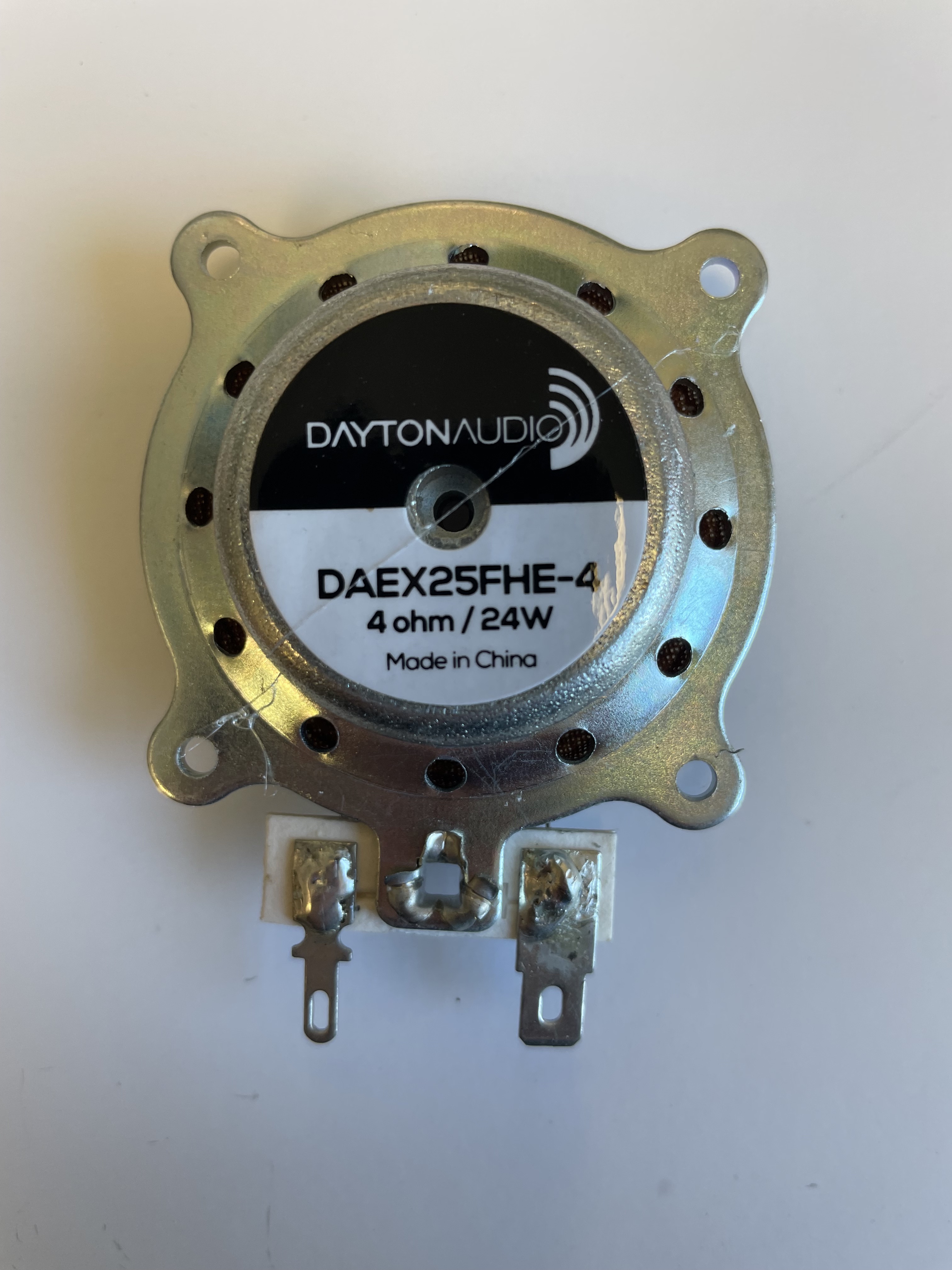}
    \label{fig:DAEX25FHE-4_Back}
  \end{subfigure}
  \hfill
  \begin{subfigure}[b]{0.32\linewidth}
    \centering
    \includegraphics[width=\linewidth]{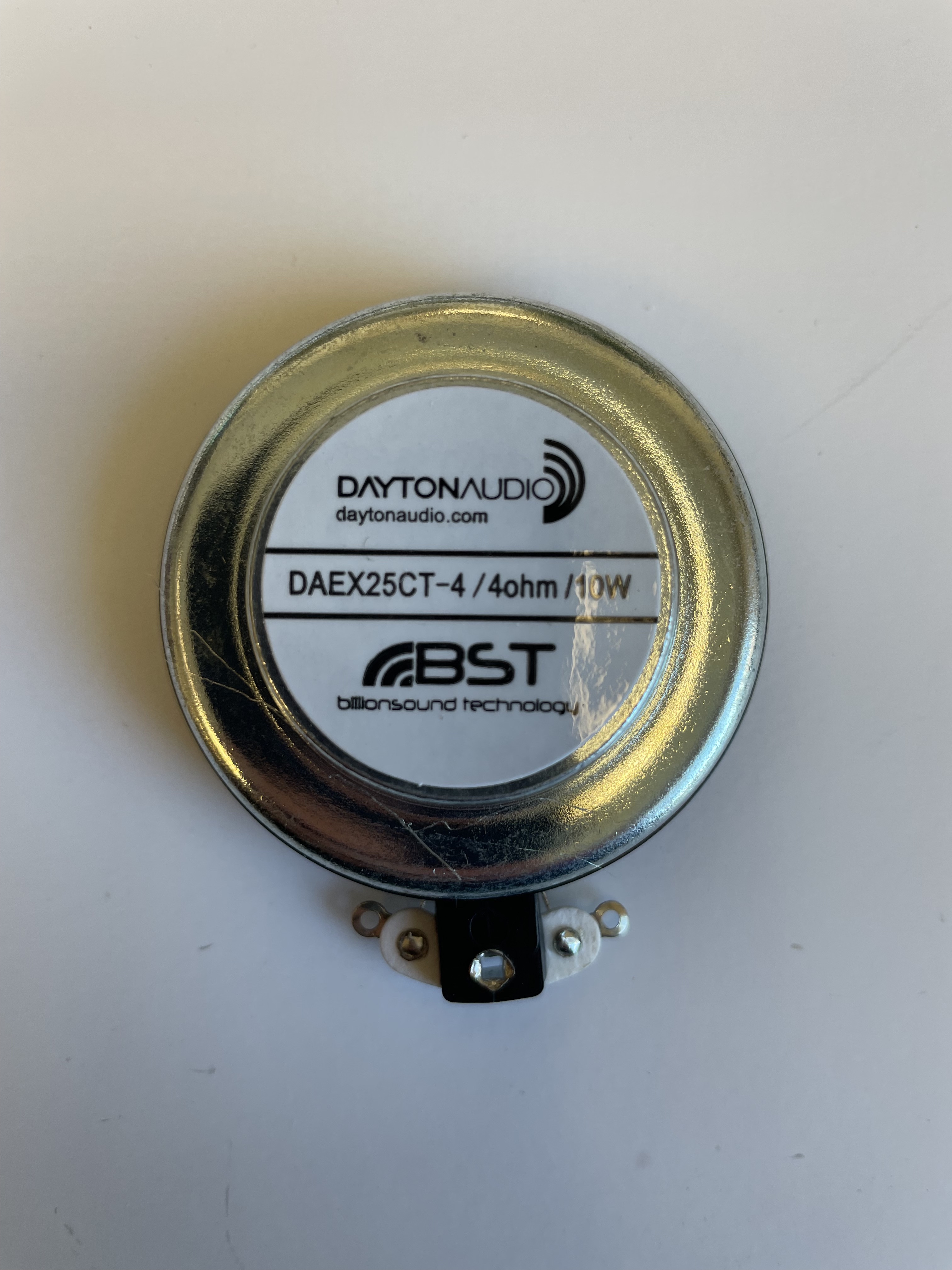}
    \label{fig:DAEX25CT-4_Back}
  \end{subfigure}
  \hfill
  \begin{subfigure}[b]{0.32\linewidth}
    \centering
    \includegraphics[width=\linewidth]{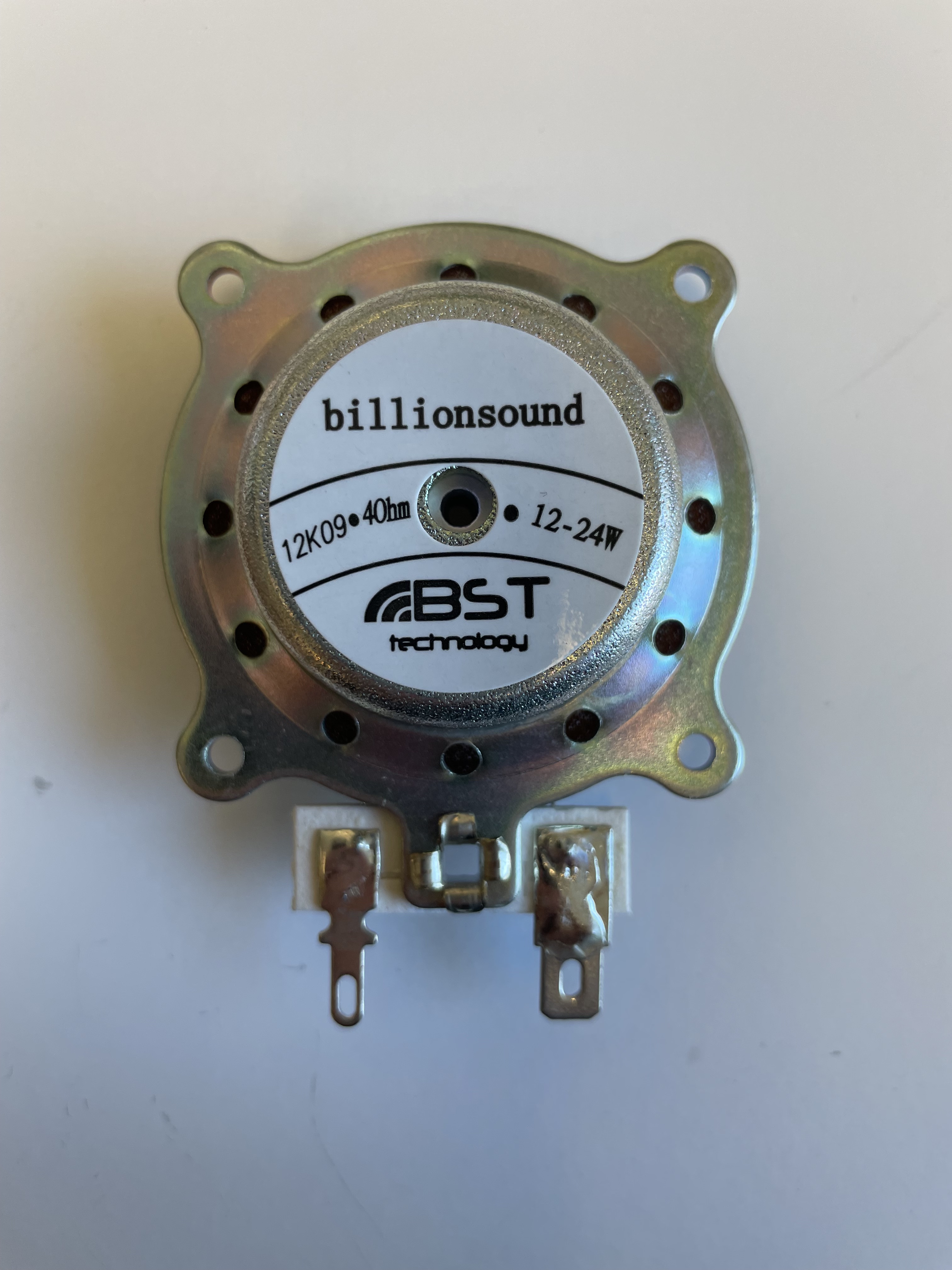}
    \label{fig:BS12K09_Back}
  \end{subfigure}
  \begin{subfigure}[b]{0.32\linewidth}
    \centering
    \includegraphics[width=\linewidth]{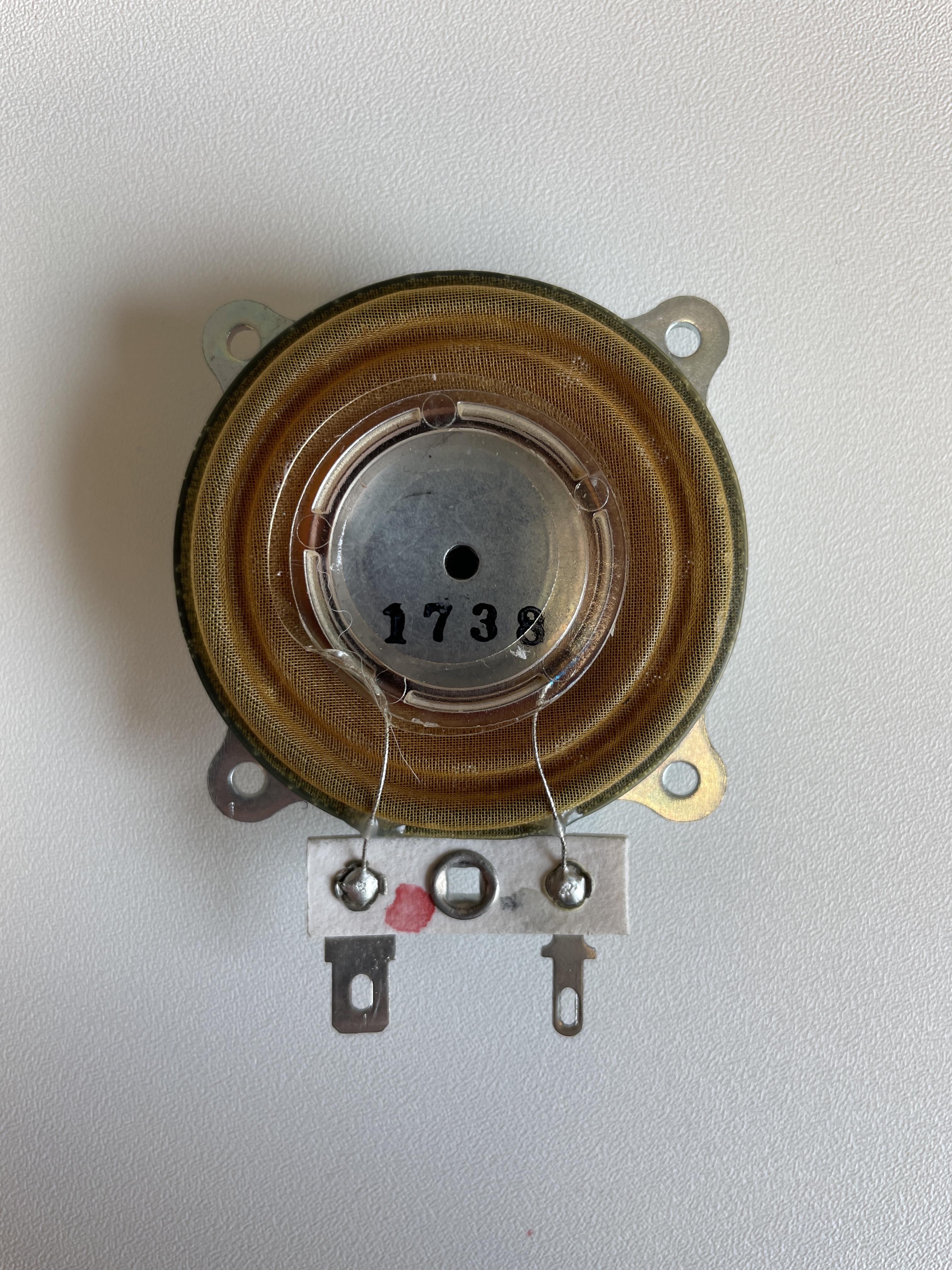}
    \caption{DAEX25FHE-4}
    \label{fig:DAEX25FHE-4_Back}
  \end{subfigure}
  \hfill
  \begin{subfigure}[b]{0.32\linewidth}
    \centering
    \includegraphics[width=\linewidth]{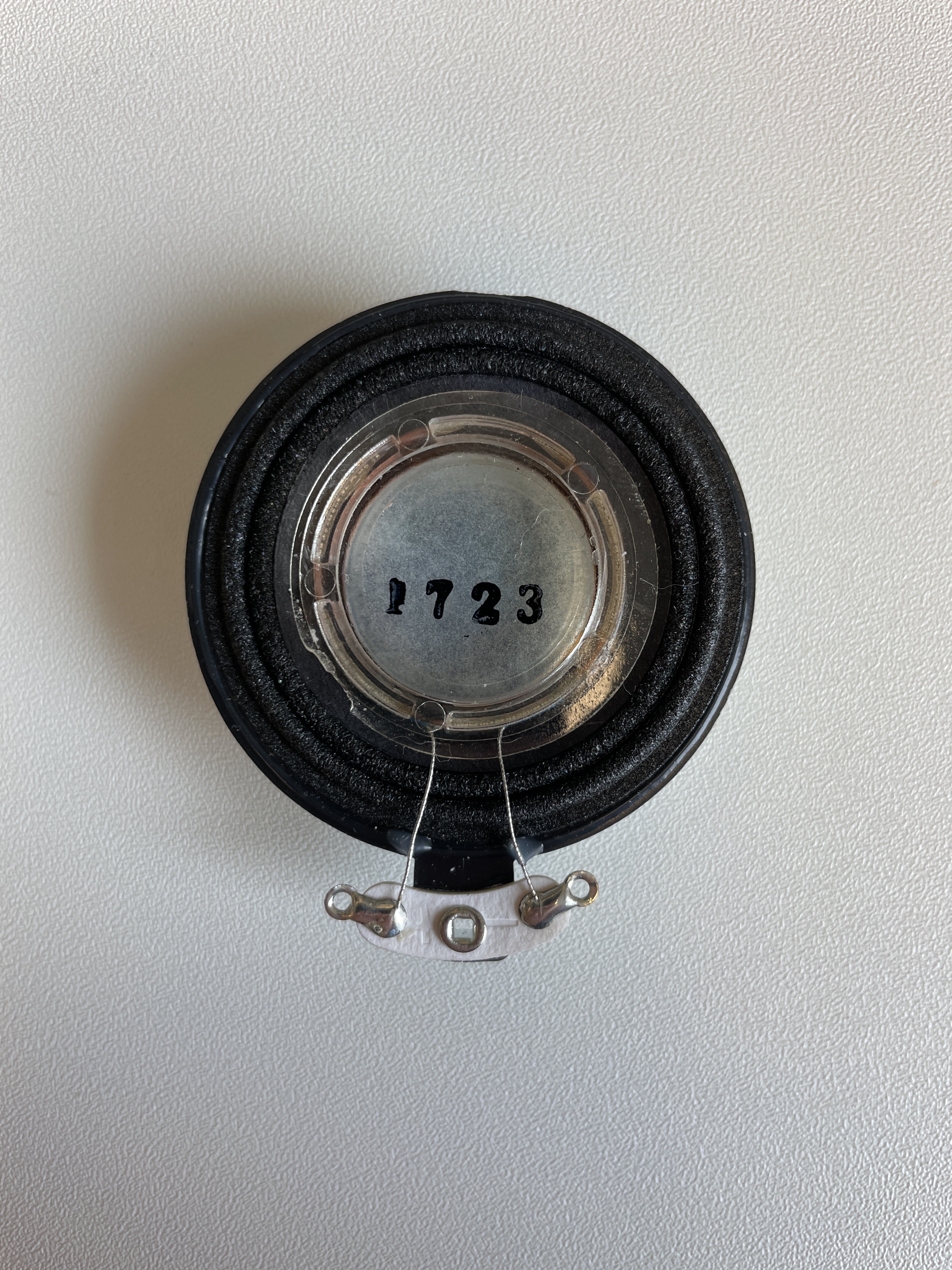}
    \caption{DAEX25CT-4}
    \label{fig:DAEX25CT-4_Back}
  \end{subfigure}
  \hfill
  \begin{subfigure}[b]{0.32\linewidth}
    \centering
    \includegraphics[width=\linewidth]{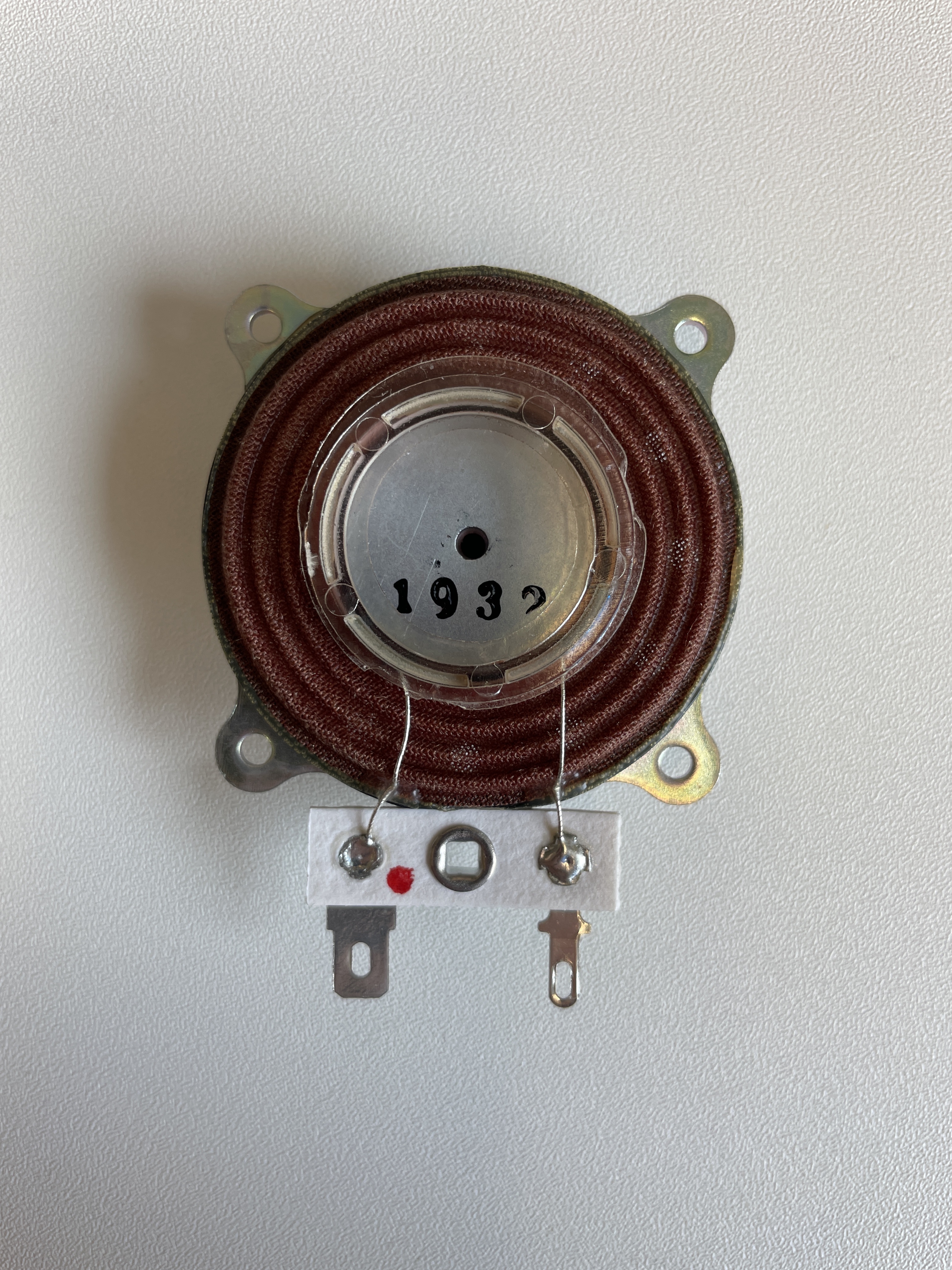}
    \caption{BS12K09}
    \label{fig:BS12K09_Back}
  \end{subfigure}
  \caption{Images of Exctiers}
\end{figure}

\begin{table*}[t]
\begin{center}
\begin{tabular}{ |c|c|c|c|c|c|c| } 
\hline
Brand & Model & Power (W) & No. Tested & Diameter (mm) & $f_0$ (Hz) & $Bl$ (Tm)\\
\hline
Dayton Audio & DAEX25FHE-4 & 24 & 10 & 27.6 & 224 & 1.54\\
Dayton Audio & DAEX25CT-4  & 10 & 10 & 27 & 306 & 3.63\\
Billionsound & BS12K09  & 24 & 5 & 27.6 & \textasciitilde240 & \textasciitilde4 \\
\hline
\end{tabular}
\caption{Manufacturer's exciter specifications and amount tested. The diameter reported is of the outer edge of the adhesive ring. Note that the datasheet for the BS12K09 is low resolution, so $f_0$ and $Bl$ are approximated. All exciters are 4$\Omega$.}
\label{table:exciterspecs}
\end{center}
\end{table*}

An Audiomatica CLIO 12 system was used to measure each exciter. To calibrate the system, the CLIO was set to generate a 1kHz sine at 0.5V\textsubscript{p}. The signal was then fed to an Emotiva BasX A2 amplifier, through the exciter, and measured with a PCB Piezotronics U352C66 accelerometer placed on the back of the exciter. The accelerometer output was sent through a PCB 481 amplifier whose output was then read by an oscilloscope and the CLIO system. On both the oscilloscope and CLIO system, the gain added was 0.5V\textsubscript{p}. No THD, IMD, or linearity distortions were found.


In a traditional loudspeaker, the surround is the component that holds the cone in place and helps guide the direction of travel of the voice coil through a single directional plane like a piston. In an exciter, the surround is made of rubber or plastic, and allows the cone to move in and out, while remaining attached to the basket that supports the cone. The behavior of the surround is similar to a non-linear spring, as the displacement rises in a speaker the surround responds with more restoring force, $F_r=R_{ms}$(x). Examples of loudspeaker surround stiffness measurements may be found in \cite{klippel2001speaker}. The non-linear action of the surround may cause distortions in the loudspeaker, which can vary depending on the construction and stiffness of the surround.

\begin{figure}[t]
\centering
\includegraphics[width = 0.75\linewidth]
{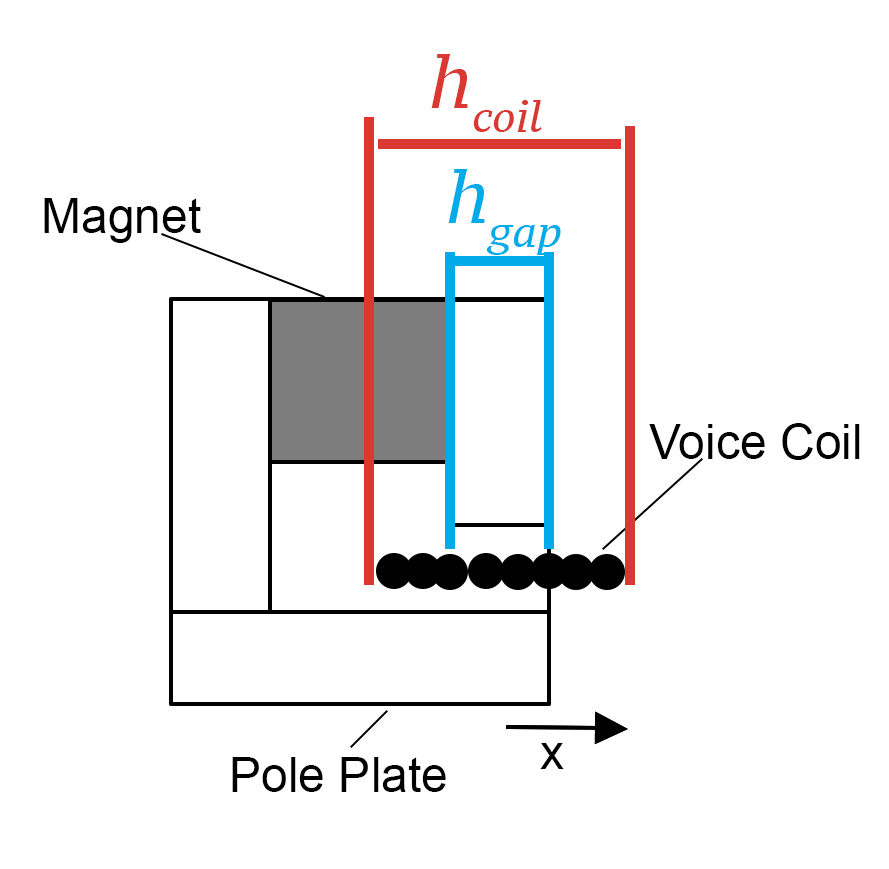}
\caption{A coil-gap configuration where the voice coil has a greater height than the gap. Reproduced from \cite{klippel2001speaker}.}
\label{Coil Gap}
\end{figure}

\begin{figure}[t]
\centering
\includegraphics[width = .75\linewidth]
{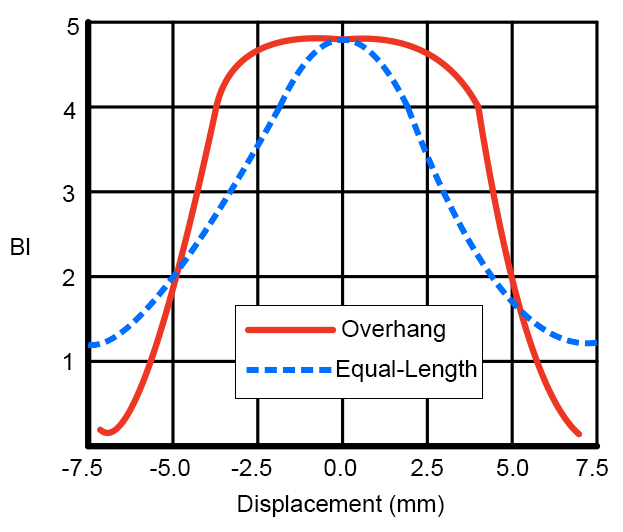}
\caption{Force Factor performance of a loudspeaker, $Bl(x)$, with an equal-length and overhang configuration. Reproduced from \cite{klippel2001speaker}.}
\label{Bl_gap_curve}
\end{figure}

The force factor, $Bl$(x), is a coupling of the electrical and mechanical components of the voice coil, where $B$ is the magnetic flux density, and $l$ is the length of the voice coil wire in the magnetic field \cite{thiele1971loudspeakers}. Since the voice coil is moving in response to the current, the length of wire in the magnetic field changes with the displacement $(x)$ of the voice coil. The coil-gap width, which is the difference between the height of the coil, $h_{coil}$, and the height of the gap, $h_{gap}$, has a great effect on the $Bl$(x) curve. The cross-section of an exciter with $h_{coil}$ wider than $h_{gap}$, known as overhang is shown in Fig.~\ref{Coil Gap} and the effect of the overhang on $Bl$ is shown in Fig. \ref{Bl_gap_curve}. If the voice coil is not equal-length to the gap and not centered on the magnet, the exciter will exhibit nonlinear distortions which can be preventable with quality controls during manufacturing.

\subsection{Mass-loaded measurements}
The resonant frequency of each exciter may be modeled as a mass-spring system with a resonant frequency $f_0$ as reported in manufacturer data sheets \cite{dickason2000loudspeaker}. However, this resonance represents the unloaded resonance of the system, where the reaction mass is coupled to free-air. It is also necessary to measure the resonances of the exciters when mass-loaded, as this represents a more realistic representation of their behavior when affixed to a panel loudspeaker. This is a critical design consideration, as from Anderson \textit{et al.} \cite{anderson2017flat-panel} the exciter resonance is coupled to the panel mode resonances, which affect the overall response of the coupled panel-exciter system if the mass-loaded resonant frequency $f'_0$ occurs near the resonant frequency of a bending mode. $f'_0$ is given by, 
\begin{equation}
\label{fspeakermounted}
f'_0  = \sqrt{\frac{R_{ms}(x)}{m}}
\end{equation}
where $R_{ms}$(x) is the stiffness of the exciter and $m$ is the inertial mass. 

\begin{figure}[t]
\centering
\includegraphics[width=0.9\linewidth]{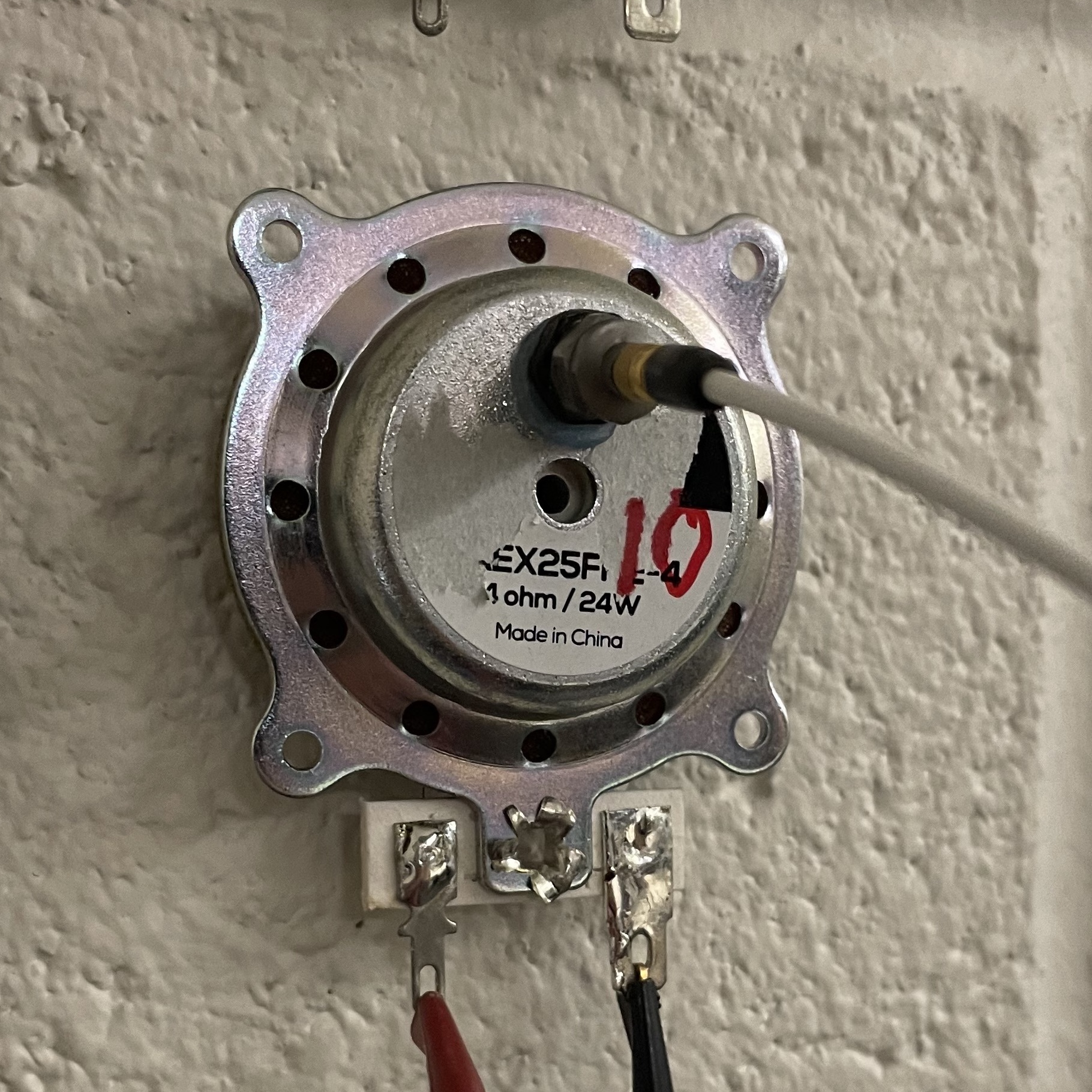}
  \caption{DAEX25FHE-4 exciter mounted to wall for mass-loaded measurements}
  \label{fig:wall_exciter0}
\end{figure}

The exciters under test were mounted on a concrete wall to simulate an infinite mass as shown in Fig.~\ref{fig:wall_exciter0} and were measured using a log chirp signal from 20Hz-20kHz, at 128kHz resolution. The mass-loaded measurements taken for each exciter included frequency response, linearity, IMD, THD, and Force vs. displacement.


\subsection{Linearity}
Loudspeakers are inherently non-linear, however, testing the linearity should give greater insight into how well each exciter is designed. The test was configured to run a 20Hz to 20kHz sweep from 0.01V to 1.5V through each exciter. The movement of the exciter was then captured by the attached accelerometer and processed through MATLAB.

\subsection{Intermodulation Distortion}
The Intermodulation Distortion (IMD) of the exciters was tested using both SMPTE \cite{SMPTE-RP120-2005} and DIN \cite{DIN45403-1} standards using a stepped voltage sweep from 0.01V to 1.5V. SMPTE IMD uses a 60Hz sine combined with a 7kHz sine at a 4:1 amplitude ratio, which reveals nonlinearities within low frequencies. The DIN IMD standard combines a 250Hz sine with an 8kHz sine at equal amplitudes. This reveals problems with higher-order distortions and nonlinearities in higher frequencies. 

\subsection{Total Harmonic Distortion}
The THD of the system and exciters used a 1kHz sine with a stepped voltage sweep from 0.01V to 1.5V. THD is typically expressed as a percentage of the energy contained in the harmonics relative to the energy contained in the fundamental.

\subsection{Force v Displacement}
Force versus displacement measurements show the linearity of the exciters and also demonstrate how well the voice coil is centered in the magnetic field.  A FSA-MSL Tester \cite{FSAMSL} was used for the measurement.  Drivers with higher $Bl$ values are more effective at converting electrical current into mechanical force. A known current (1A) is applied to the exciter, and the Tester measures the force exerted by the voice coil. However, the spider also exerts its own force on the system. The resting position of the voice coil is determined by zeroing the force when zero current applied. The spider acts like a spring, and exerts a force that is a function of displacement as the voice coil moves in or out. The force is measured at each position of the voice coil without current, and the change in force is reported when a 1A DC signal is run through the voice coil.

\section{Results}

\subsection{Mass-Loaded Frequency Response}

\begin{figure}[t]
\includegraphics[width= .9\linewidth]{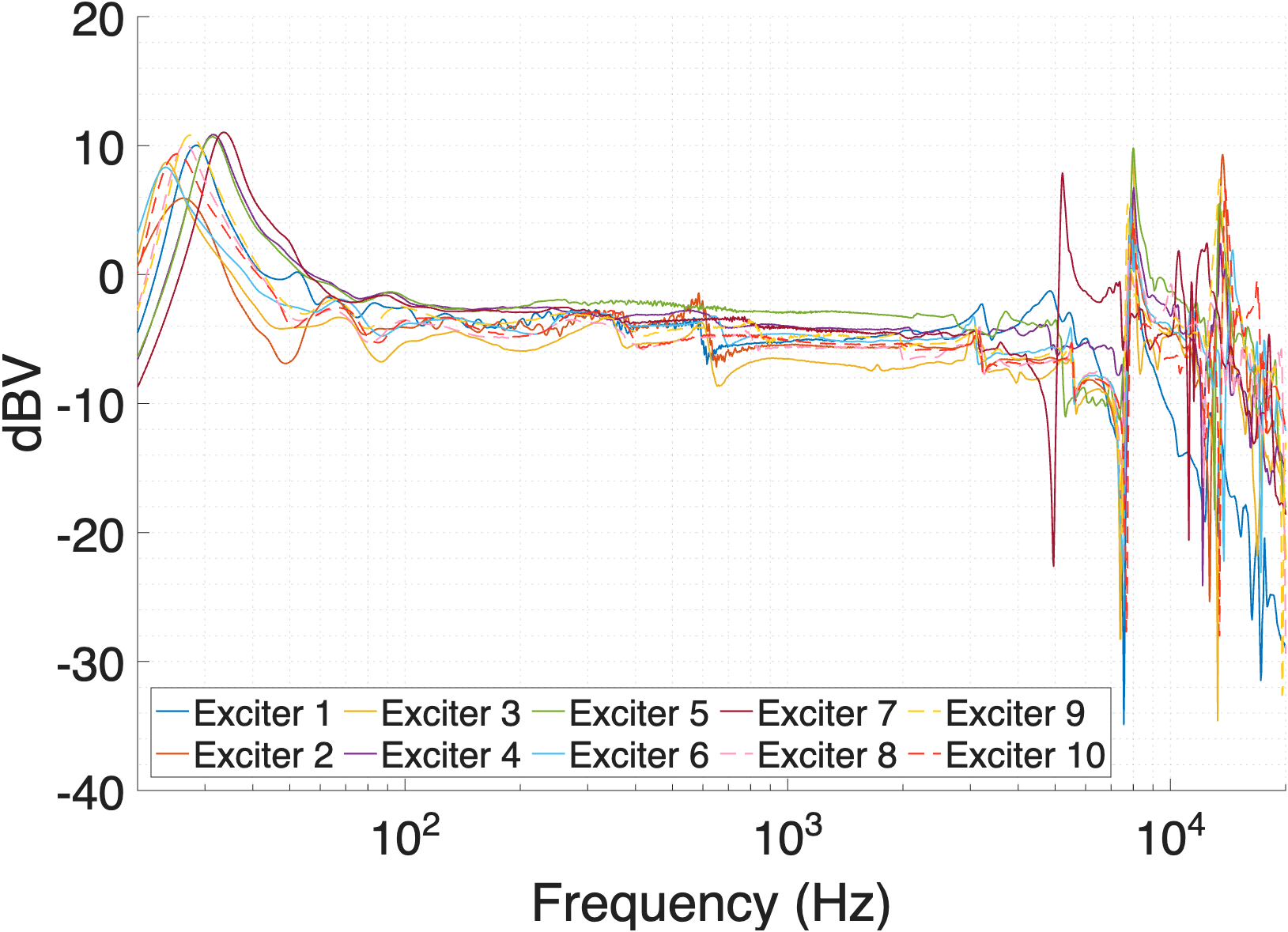}
\caption{Frequency Response of DAEX25FHE-4, $f$=20Hz-20kHz}
\label{fig:DAEX25FHE_4_Full}
\end{figure}

\begin{figure}[t]
\centering
\includegraphics[width= .9\linewidth]{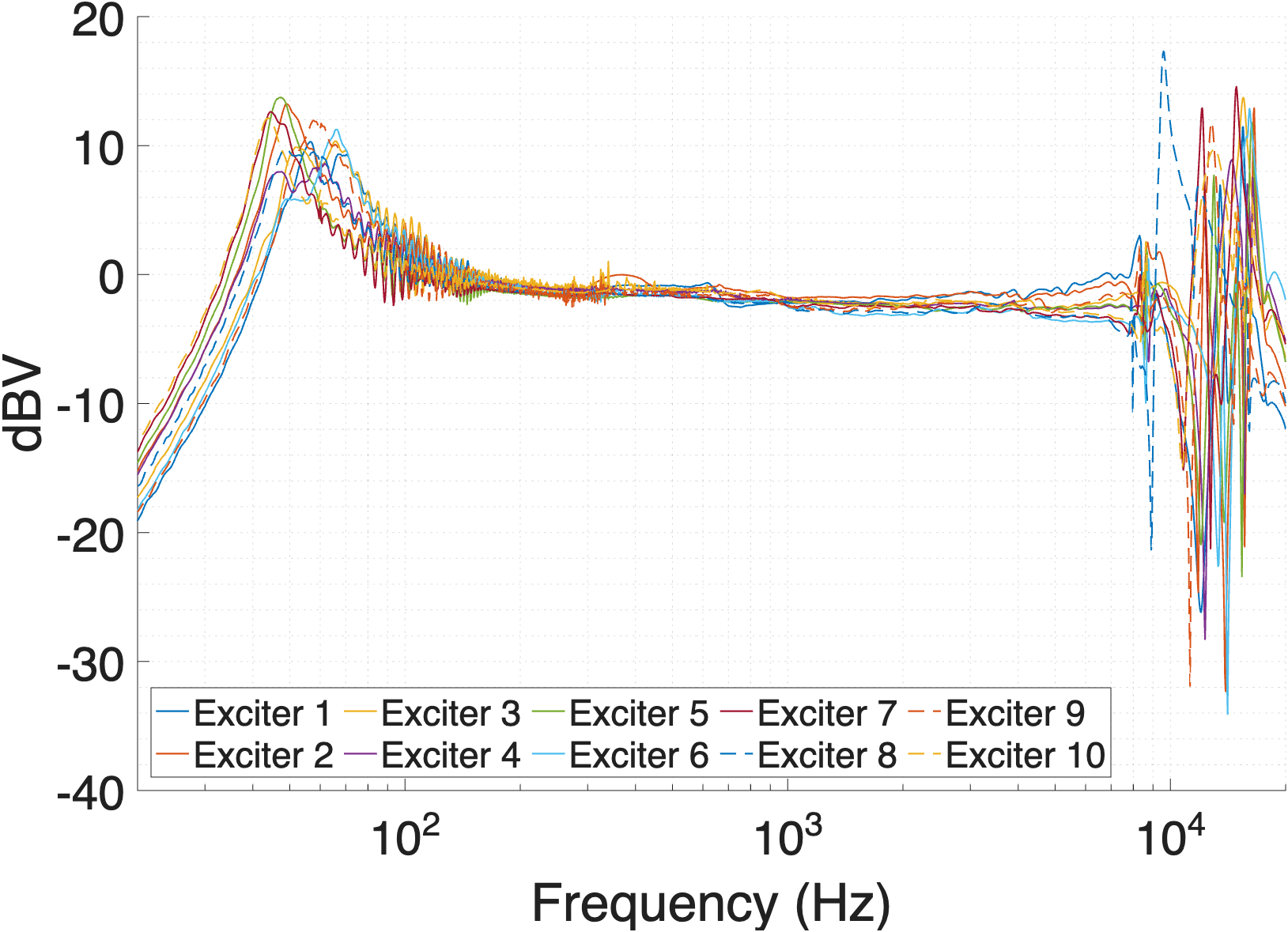}
\caption{Frequency Response of DAEX25CT-4, $f$=20Hz-20kHz}
\label{fig:DAEX25CT_4_FreqFull}
\end{figure}

\begin{figure}[t]
\centering
\includegraphics[width= .9\linewidth]{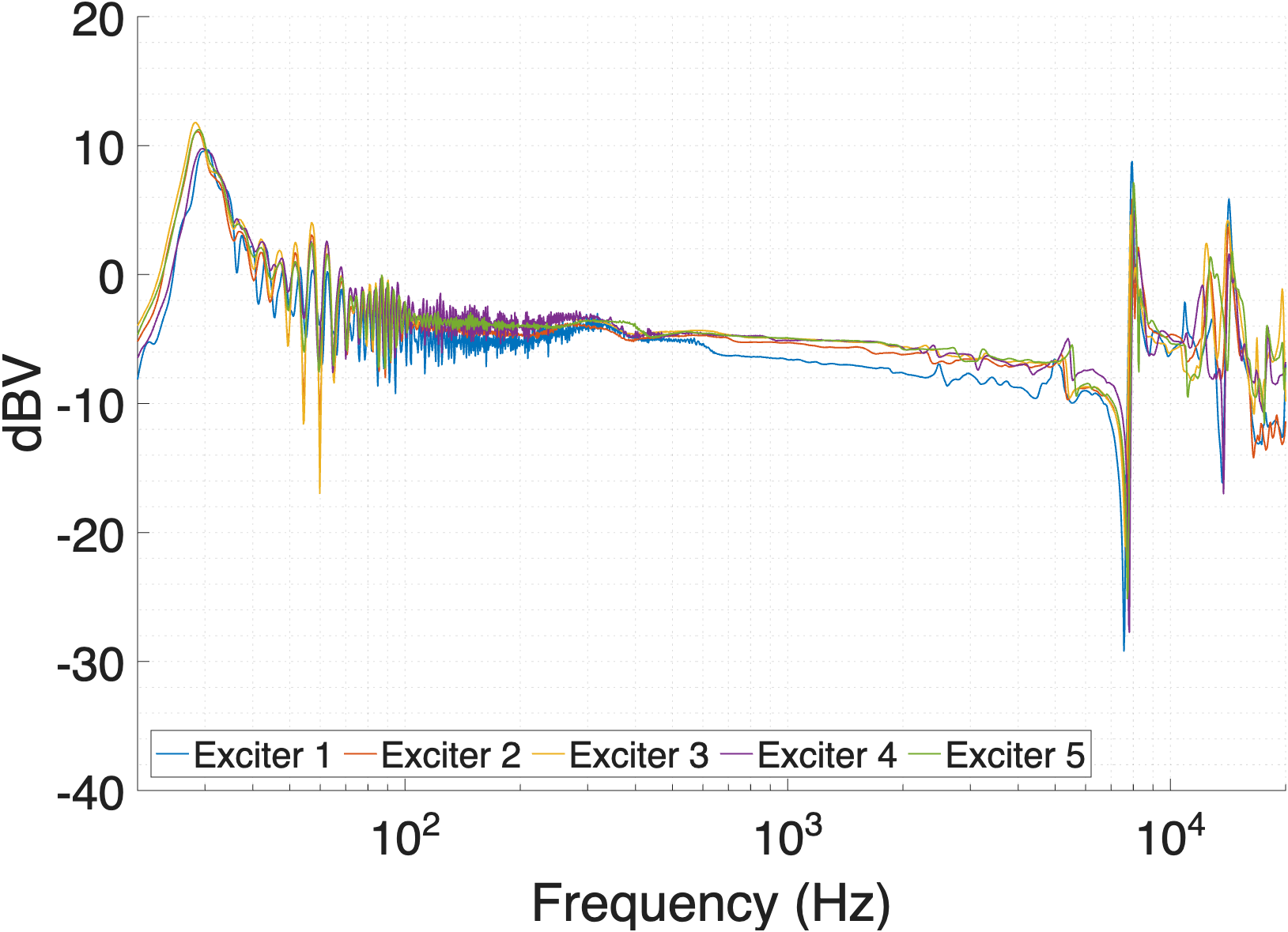}
\caption{Frequency Response of BS12K09, $f$ = 20Hz-20kHz}
\label{fig:BS12K09_FreqFull}
\end{figure}

The broadband frequency response of each exciter can be seen in Figs.~\ref{fig:DAEX25FHE_4_Full}-\ref{fig:BS12K09_FreqFull}. Each exciter exhibits extreme nonlinearities as they near 10kHz. These exciters are not effective at reproducing high frequencies where the wavelength on the panel is smaller than the diameter of the ring used to adhere the exciter to the panel. It is best practice to avoid using these larger exciters for high frequencies, and instead employ them in conjunction with smaller exciters in a crossover network scheme \cite{MXN}.

\begin{figure}[!h]
\centering
\includegraphics[width= .9\linewidth]{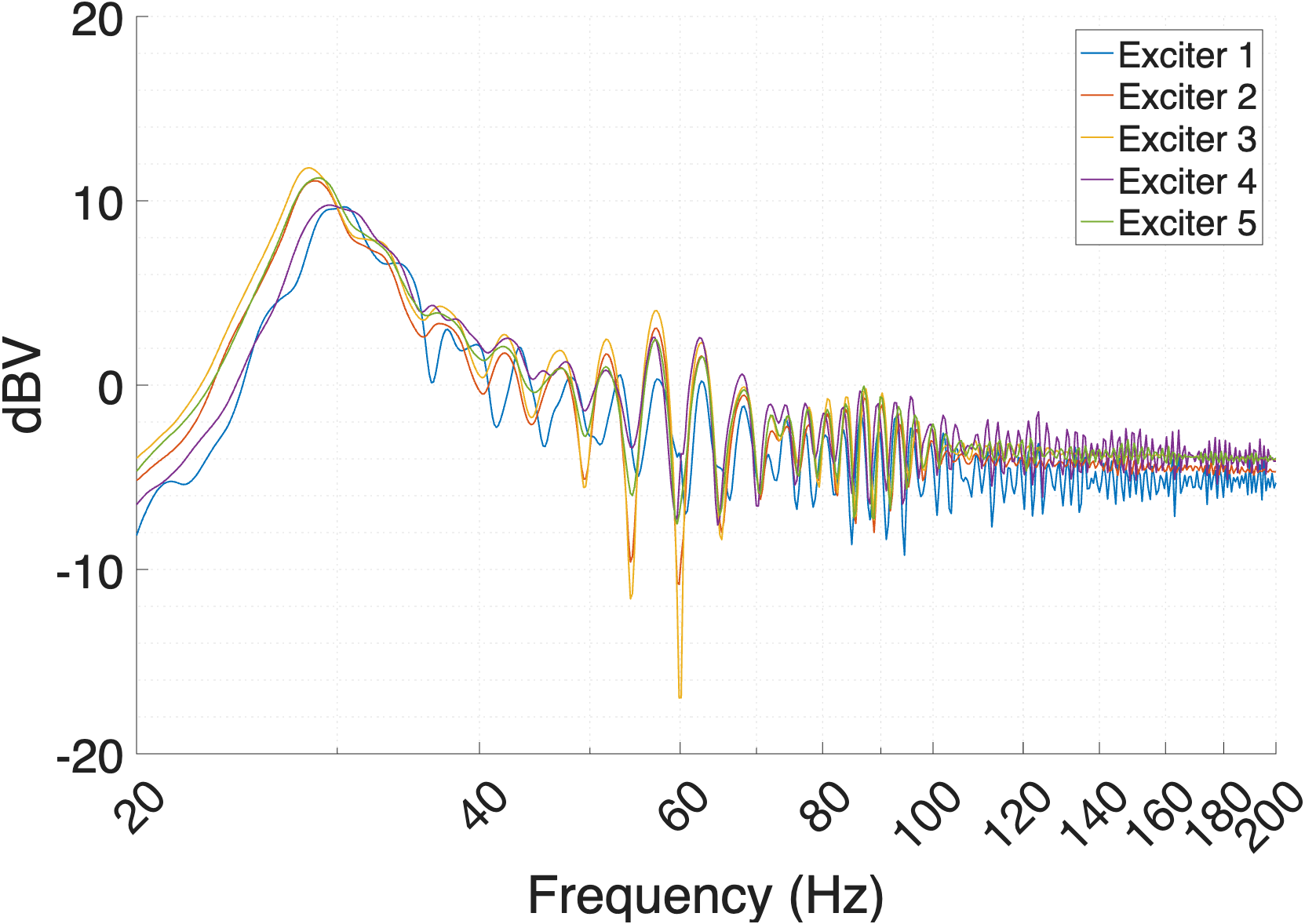}
\caption{$f'_0$ of BS12K09, $f$=20Hz-200Hz}
\label{fig:BS12K09_FreqZoom}
\end{figure}

In Fig.~\ref{fig:DAEX25FHE_4_Full}, a phase shift and dip in response occurs around 600Hz for the DAEX25FHE-4. The cause of this is suspected to be the stiffness of the suspension $K$, where stiff suspensions reduce the effectiveness of transferring higher frequencies to the baffle. In contrast, the DAEX25CT-4 has a more predictable response when compared to the DAEX25FHE-4, as seen in Fig.~\ref{fig:DAEX25CT_4_FreqFull}. The BS12K09 has a worrying ringing beginning around 80Hz which can be seen in Fig.~\ref{fig:BS12K09_FreqFull} and more closely in Fig.~\ref{fig:BS12K09_FreqZoom}.

\begin{figure}[!h]
\centering
\includegraphics[width= .9\linewidth]{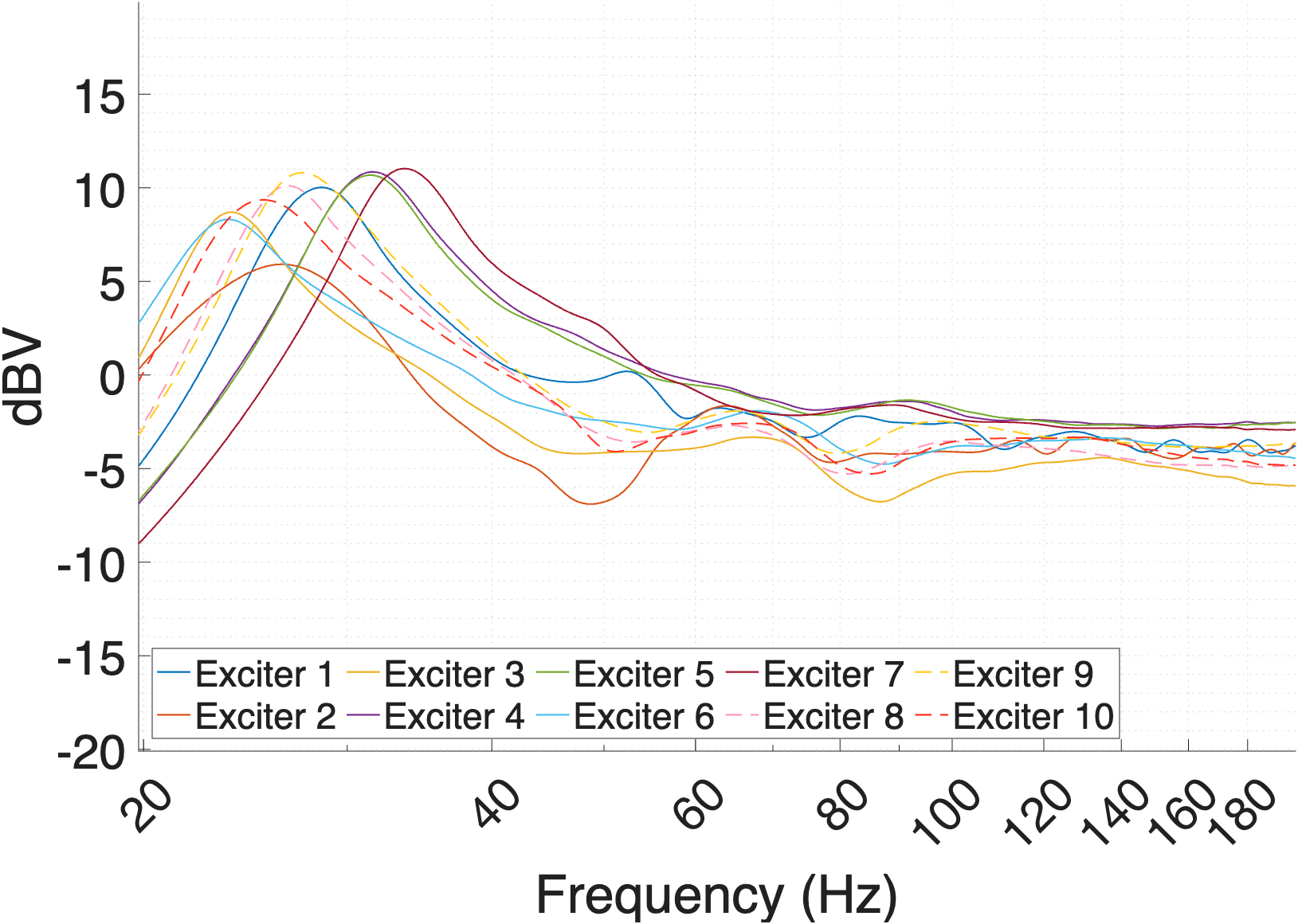}
\caption{$f'_0$  of DAEX25FHE-4, $f$ = 20Hz-200Hz}
\label{fig:DAEX25FHE_4_FreqZoom}
\end{figure}

\begin{figure}[!h]
\centering
\includegraphics[width= .9\linewidth]{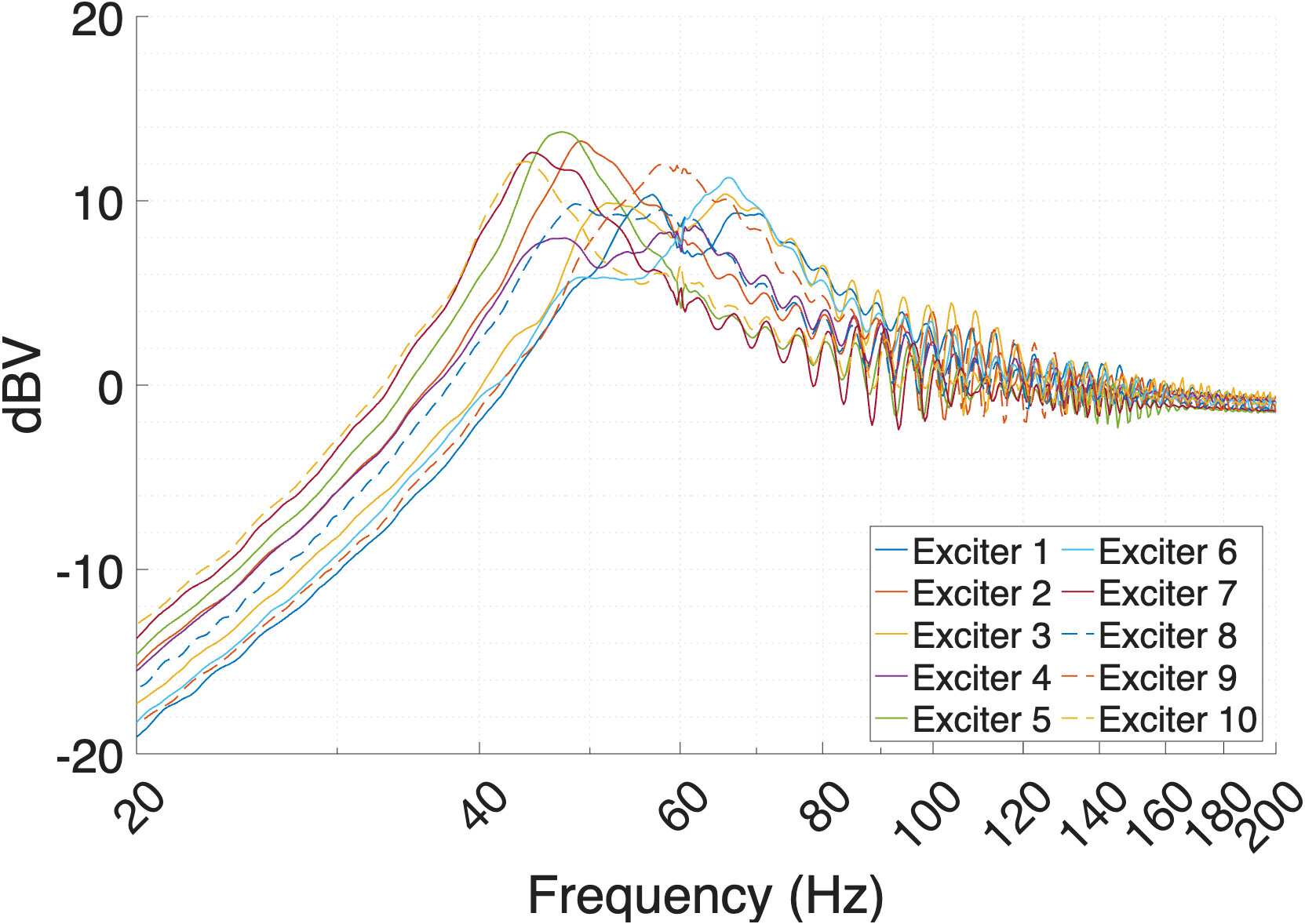}
\caption{$f'_0$ of DAEX25CT-4, $f$ = 20Hz-200Hz}
\label{fig:DAEX25CT_4_FreqZoom}
\end{figure}

Figs.~\ref{fig:DAEX25FHE_4_FreqZoom}-\ref{fig:BS12K09_FreqZoom} show the frequency response with the X-axis limited from 20Hz-200Hz, emphasizing the $f'_0$ variation in each tested exciter. The ranges of $f'_0$ for each exciter are shown in Table~\ref{table:massloaded_specs}. Concerns arise with the resonances that occur after $f'_0$, particularly in Fig.~\ref{fig:DAEX25CT_4_FreqZoom} and Fig.~\ref{fig:BS12K09_FreqZoom}. It is suspected that the suspension is not stiff enough to supply the exciter with a strong restoring force. 

\begin{table*}
\begin{center}
\begin{tabular}{ |c|c|c|c| } 
\hline
Brand & Model & $f'_0$ Range (Hz) & Mean $f'0$ (Hz)\\
\hline
Dayton Audio & DAEX25FHE-4 & 23-33 & 27.5\\
Dayton Audio & DAEX25CT-4  & 43-66 & 54.5\\
Billionsound & BS12K09  & 28-30 & 29\\
\hline
\end{tabular}

\caption{Exciter $f'_0$ variation under mass load}
\label{table:massloaded_specs}
\end{center}
\end{table*}

\subsection{Linearity}
\begin{figure}[t]
\centering
\includegraphics[width= .9\linewidth]{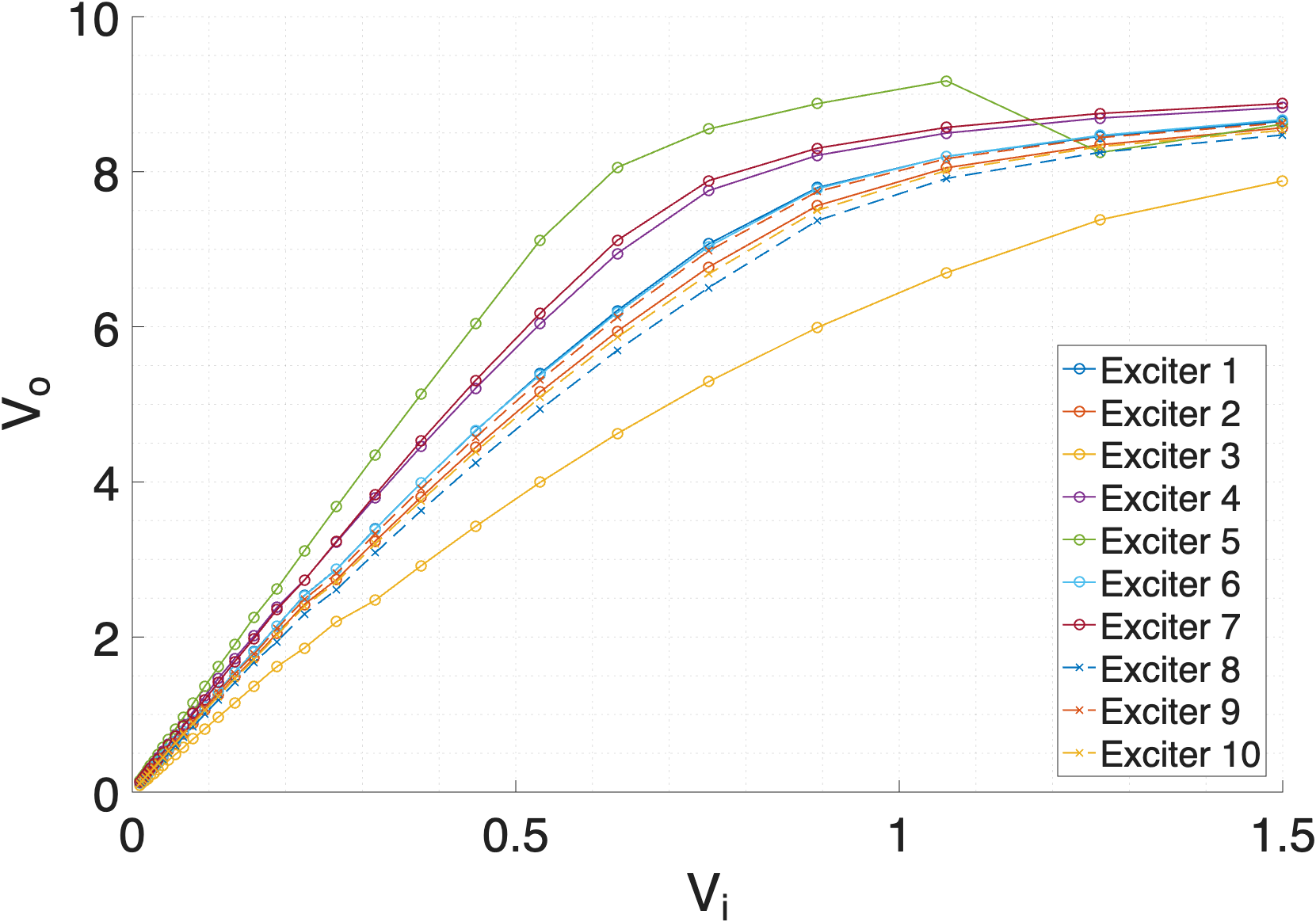}
\caption{Linearity of DAEX25FHE-4}
\label{fig:DAEX25FHE_4_Lin}
\end{figure}

\begin{figure}[t]
\centering
\includegraphics[width= .9\linewidth]{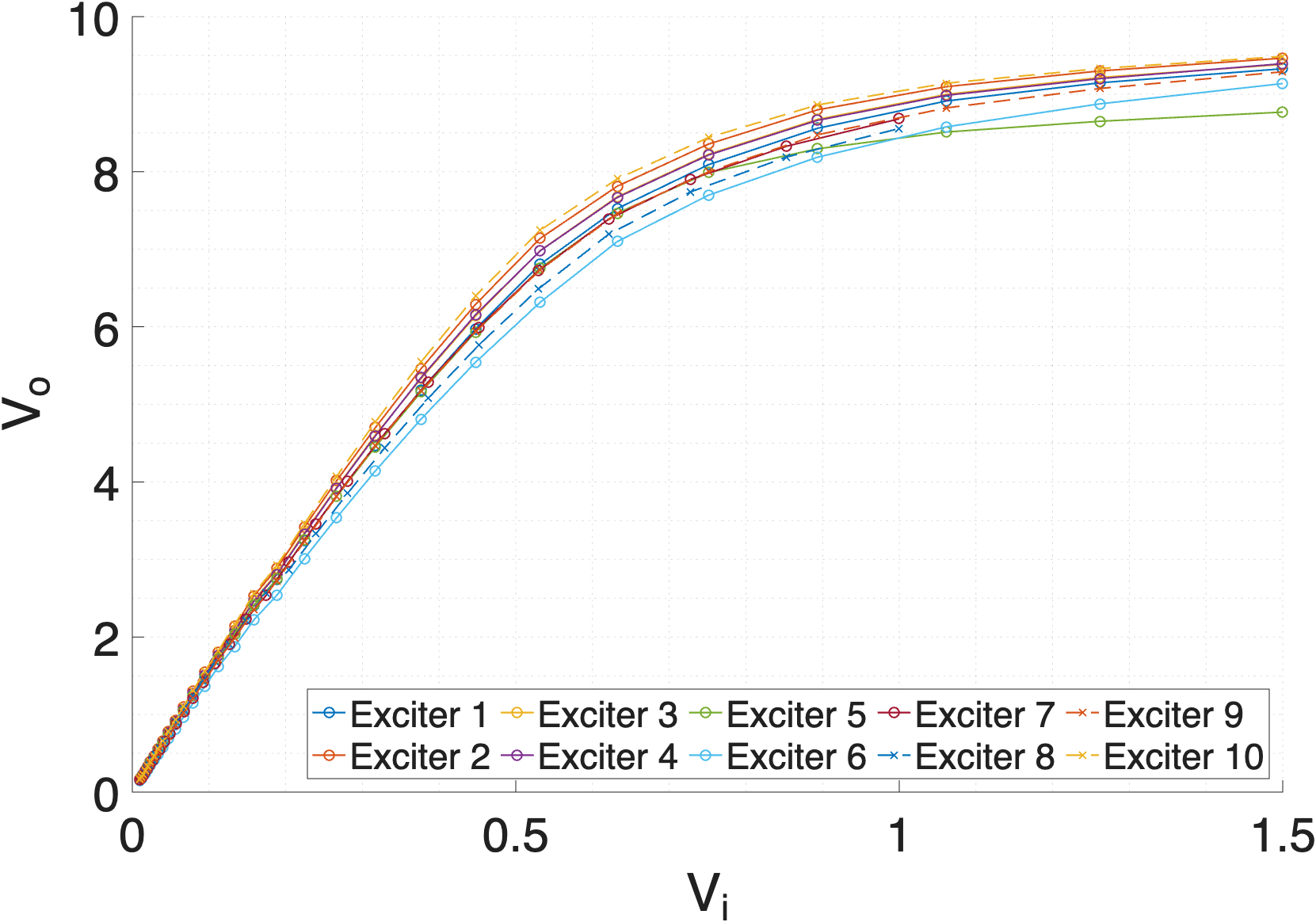}
\caption{Linearity of DAEX25CT-4}
\label{fig:DAEX25CT_4_Lin}
\end{figure}

\begin{figure}[t]
\centering
\includegraphics[width= .9\linewidth]{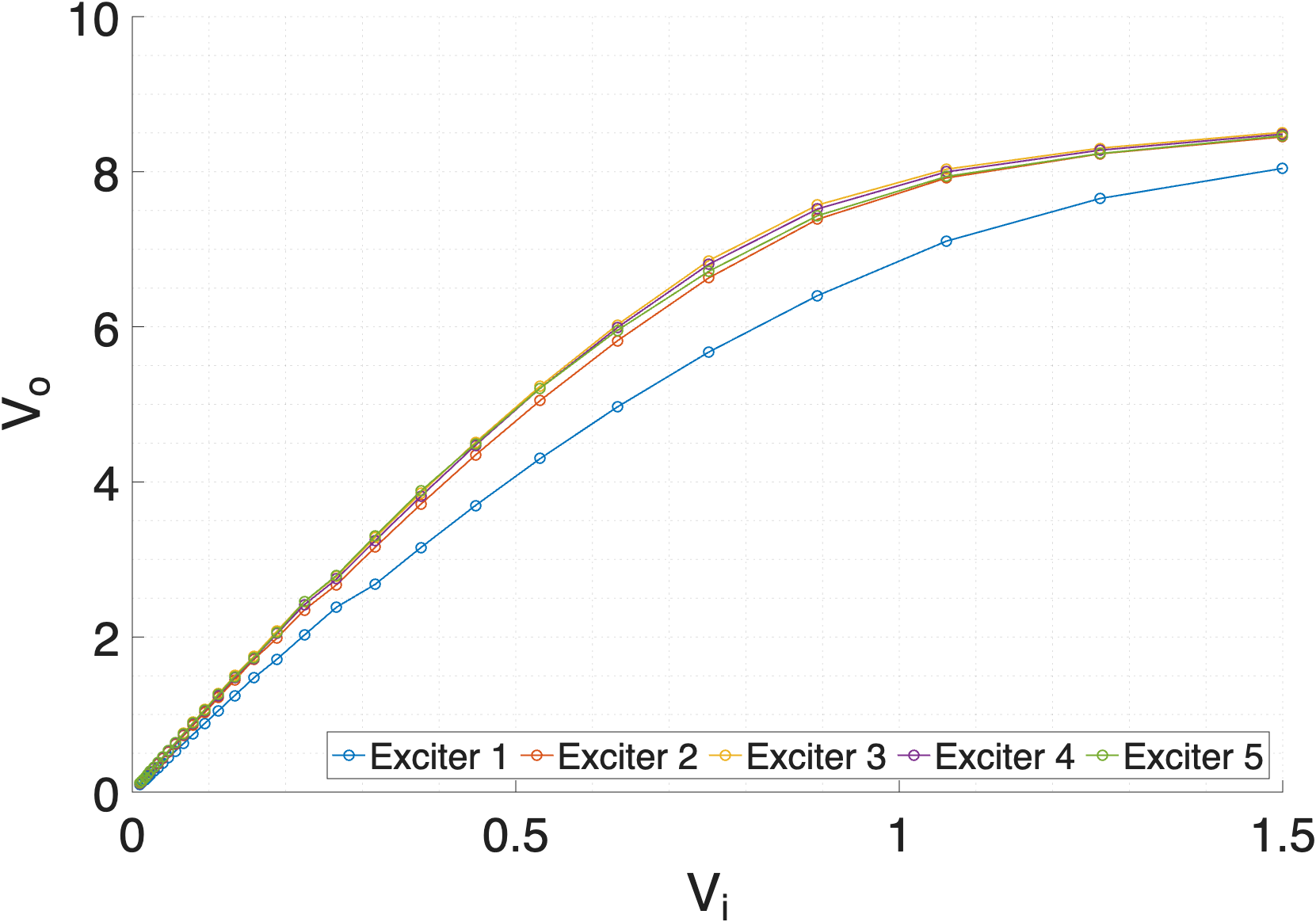}
\caption{Linearity of BS12K09}
\label{fig:BS12K09_Lin1}
\end{figure}

The linearities of the exciters are shown in Figs.~\ref{fig:DAEX25FHE_4_Lin}-\ref{fig:BS12K09_Lin1}. The DAEX25FHE-4 (Fig.~\ref{fig:DAEX25FHE_4_Lin}) exhibited a wide range of non-linearity. Assuming the construction of the cone, magnet, and surround are consistent, inconsistencies in the length of the voice coil are suspected to cause this issue. The DAEX25CT-4 (Fig.~\ref{fig:DAEX25CT_4_Lin}) had the most consistent linearity of the three exciters tested. The high linearity correlation between the exciters suggests that this model may be more easily matched between units for use in FPLs. The linearity of the BS12K09 (Fig. \ref{fig:BS12K09_Lin1}) was also fairly consistent, with one outlier in the five units tested.

\subsection{SMPTE Intermodualtion Distortion}

\begin{figure}[!h]
\centering
\includegraphics[width= .9\linewidth]{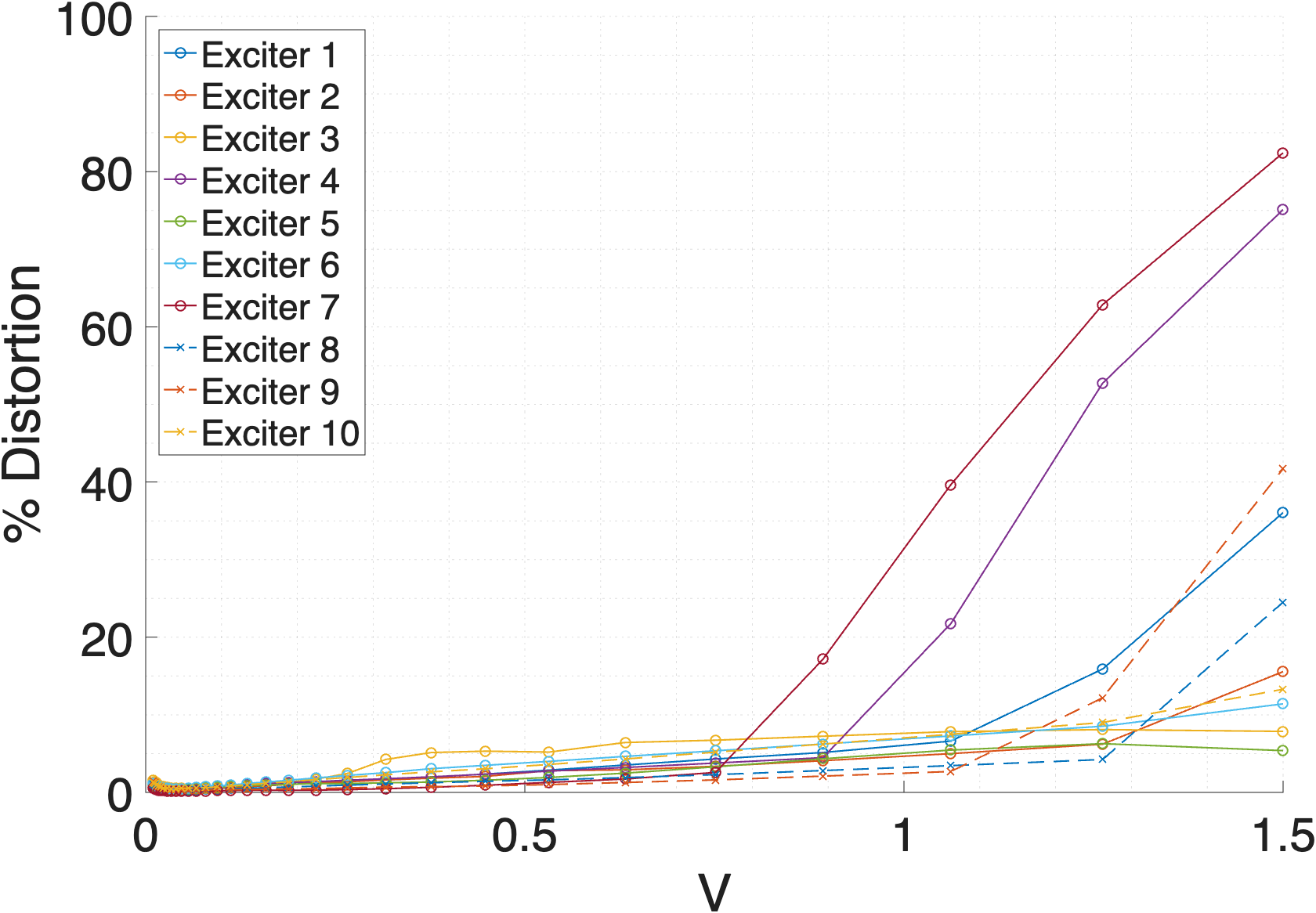}
\caption{SMPTE IMD of DAEX25FHE-4}
\label{fig:DAEX25FHE_4_IMD_SMPTE}
\end{figure}

\begin{figure}[!h]
\centering
\includegraphics[width= .9\linewidth]{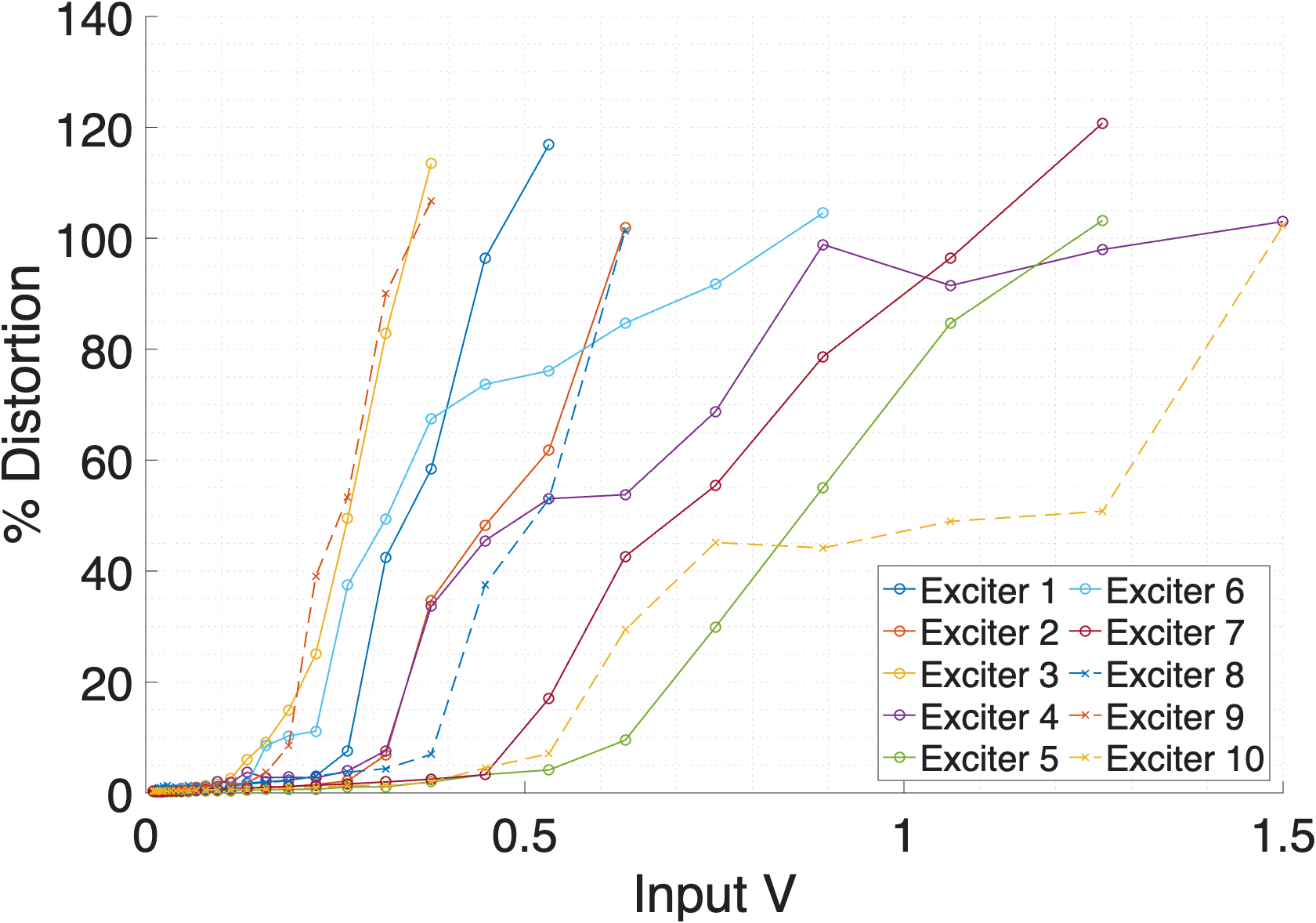}
\caption{SMPTE IMD of DAEX25CT-4}
\label{fig:DAEX25CT_4_IMD_SMPTE}
\end{figure}

\begin{figure}[!h]
\centering
\includegraphics[width= .9\linewidth]{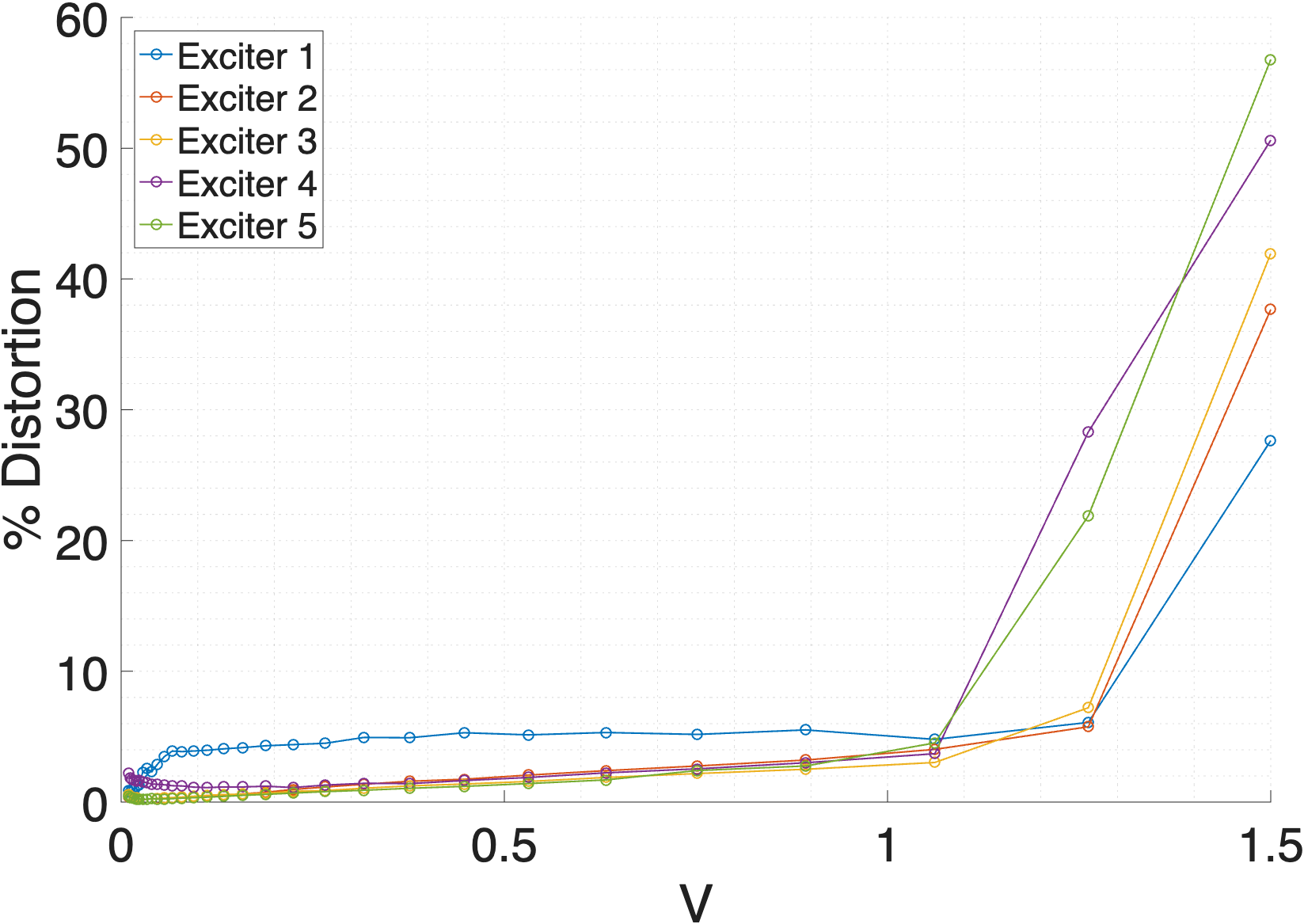}
\caption{SMPTE IMD of BS12K09}
\label{fig:BS12K09_IMD_SMPTE}
\end{figure}

The SMPTE IMD for each exciter is shown in Figs.~\ref{fig:DAEX25FHE_4_IMD_SMPTE}-\ref{fig:BS12K09_IMD_SMPTE}. Generally, if driven at low voltages ($<0.5V$) the exciters perform with low distortion with the exception of the DAEX25CT-4 (Fig.~\ref{fig:DAEX25CT_4_IMD_SMPTE}). The $f_1$ for the SMPTE measurement is 60Hz, which is fairly close to the mean $f'_0$ of 54.5Hz for this exciter. With the resonance of the exciter so close to the SMPTE $f_1$, combined with the ringing above $f'_0$ seen in Fig.~\ref{fig:DAEX25CT_4_FreqZoom}, the SMPTE IMD of this exciter is noticeable even at low input voltages.

\subsection{DIN Intermodualtion Distortion}
\begin{figure}[!h]
\centering
\includegraphics[width= .9\linewidth]{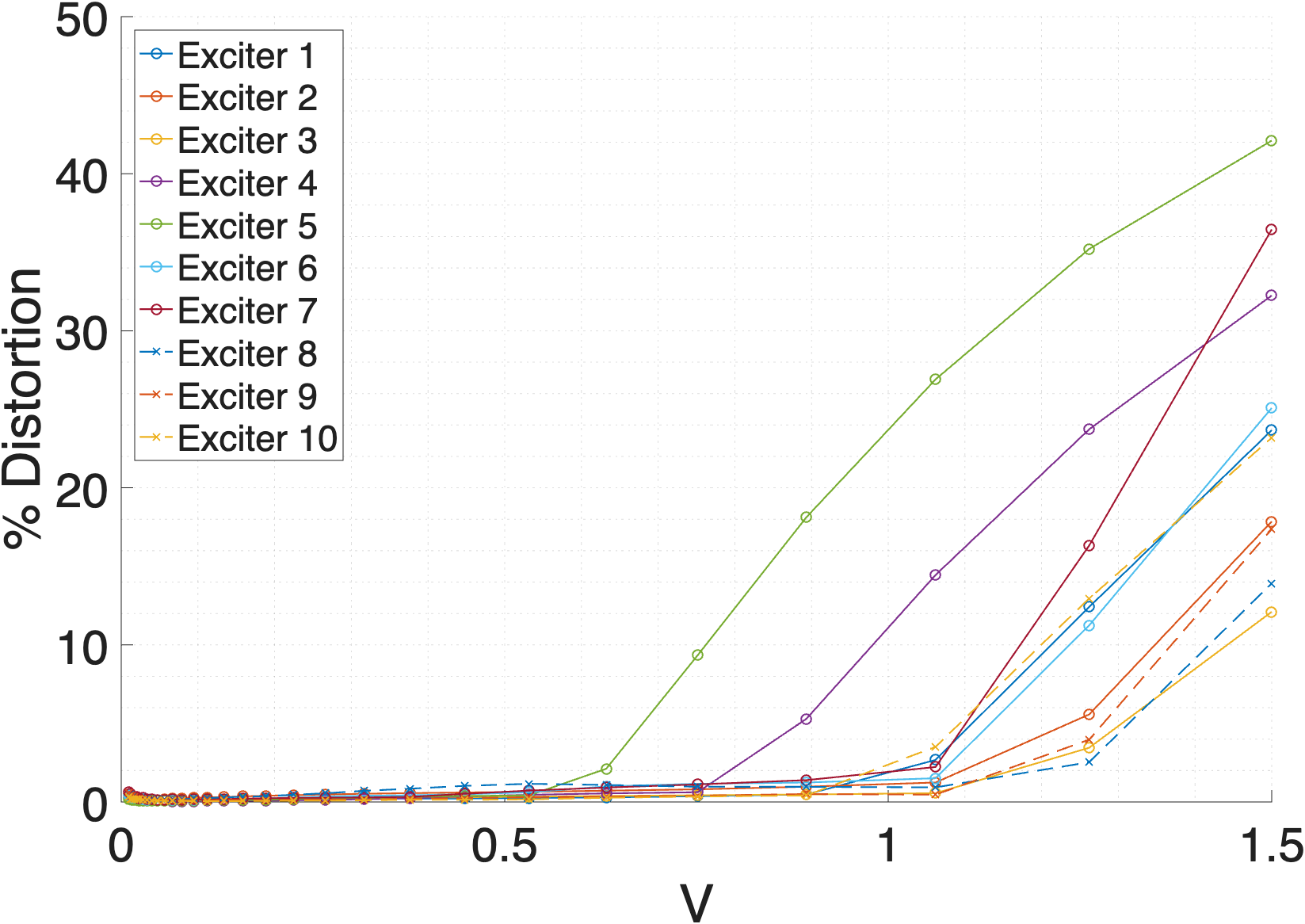}
\caption{DIN IMD of DAEX25FHE-4}
\label{fig:DAEX25FHE_4_IMD_DIN}
\end{figure}

\begin{figure}[!h]
\centering
\includegraphics[width= .9\linewidth]{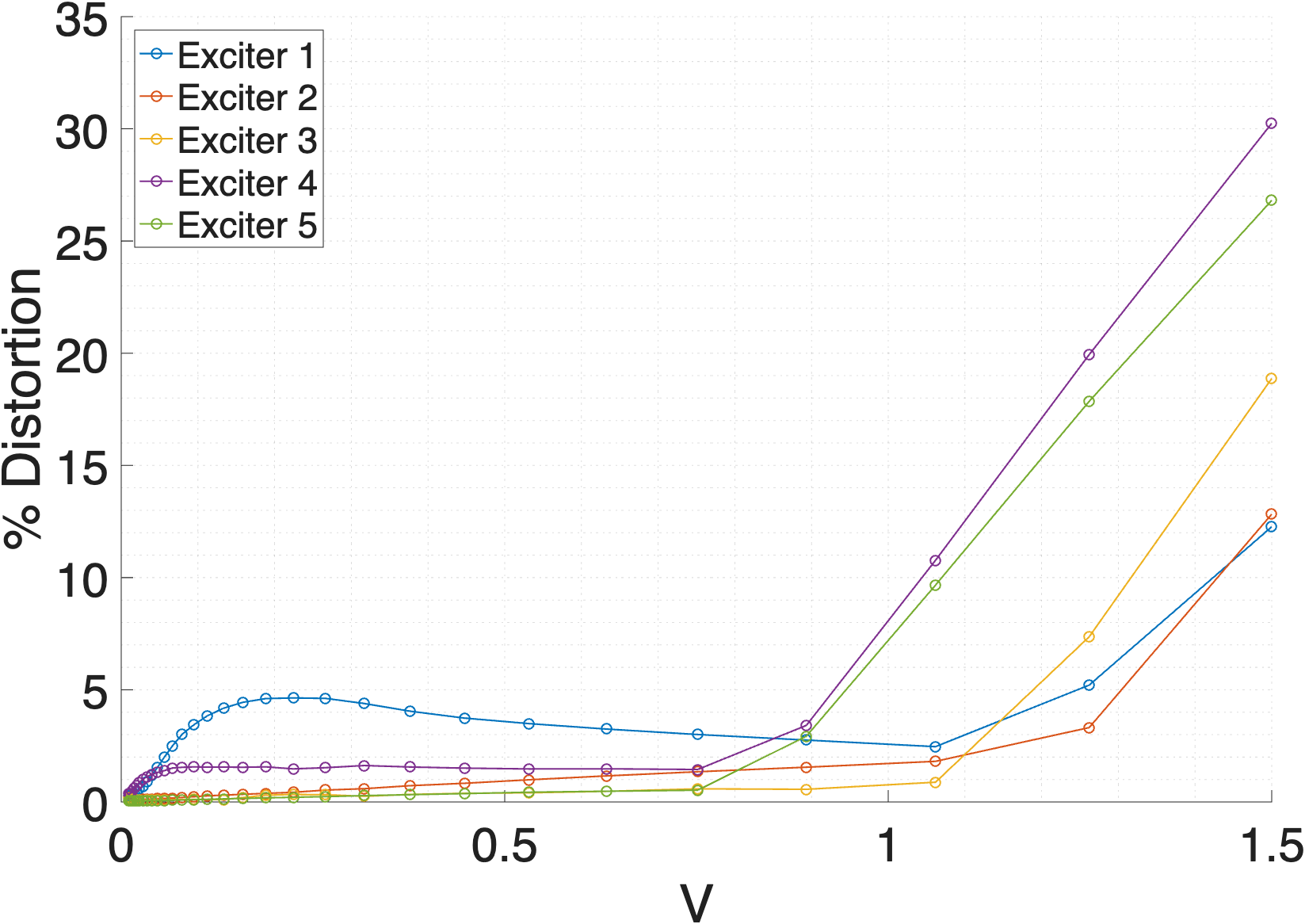}
\caption{DIN IMD of BS12K09}
\label{fig:BS12K09_IMD_DIN}
\end{figure}

Comparatively, the DIN IMD measurement is more consistent for these exciters accross the board. Since the DIN standard calls for the $f_1$ at 250Hz, the distortions seen in Figs.~\ref{fig:DAEX25FHE_4_IMD_SMPTE}-\ref{fig:BS12K09_IMD_SMPTE} are not as severe. Again, each exciter performed differently due to construction variations and lack of quality control. As an example of how these measurements relate, we can clearly see how Exciter 1 of the BS12K09 has poor performance in DIN IMD (Fig.\ref{fig:BS12K09_IMD_DIN}), which is also seen in SMPTE IMD (Fig.\ref{fig:BS12K09_IMD_SMPTE}) and Linearity (Fig.\ref{fig:BS12K09_Lin1}).

\begin{figure}[!h]
\centering
\includegraphics[width= .9\linewidth]{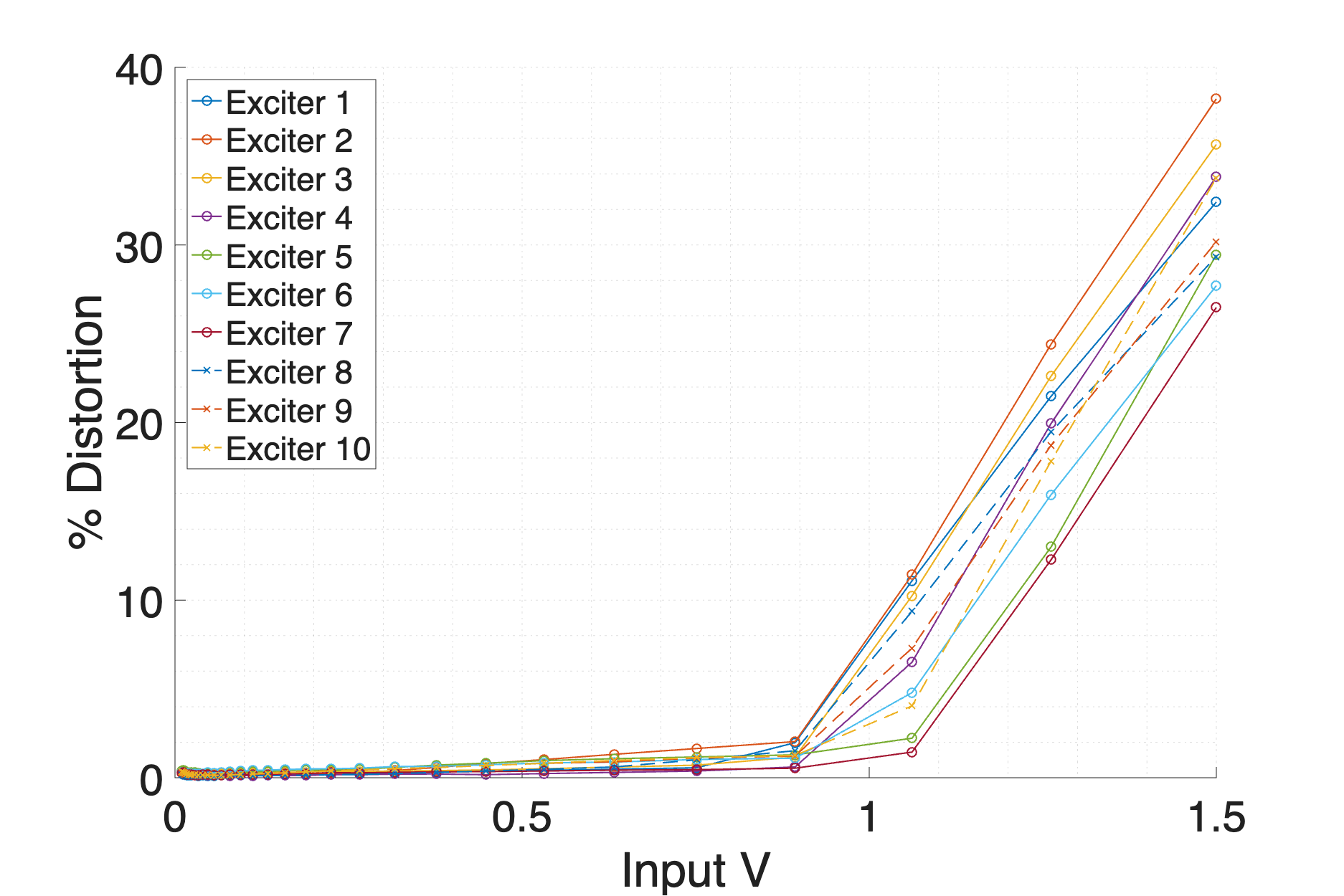}
\caption{DIN IMD of DAEX25CT-4}
\label{fig:DAEX25CT_4_IMD_DIN}
\end{figure}

\subsection{Total Harmonic Distortion}
\begin{figure}[!h]
\centering
\includegraphics[width= .9\linewidth]{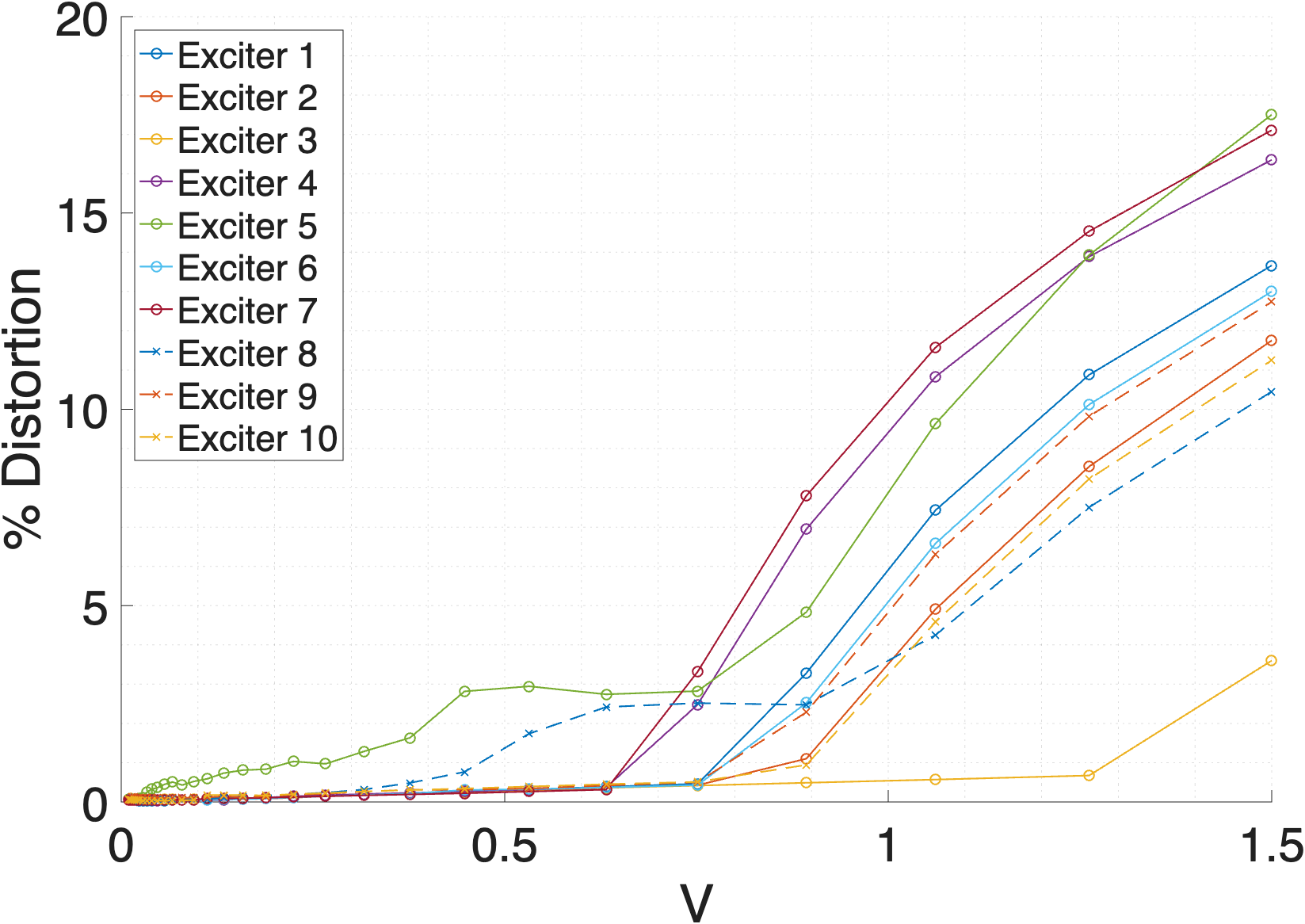}
\caption{THD of DAEX25FHE-4}
\label{fig:DAEX25FHE_4_THD}
\end{figure}

\begin{figure}[h]
\centering
\includegraphics[width= .9\linewidth]{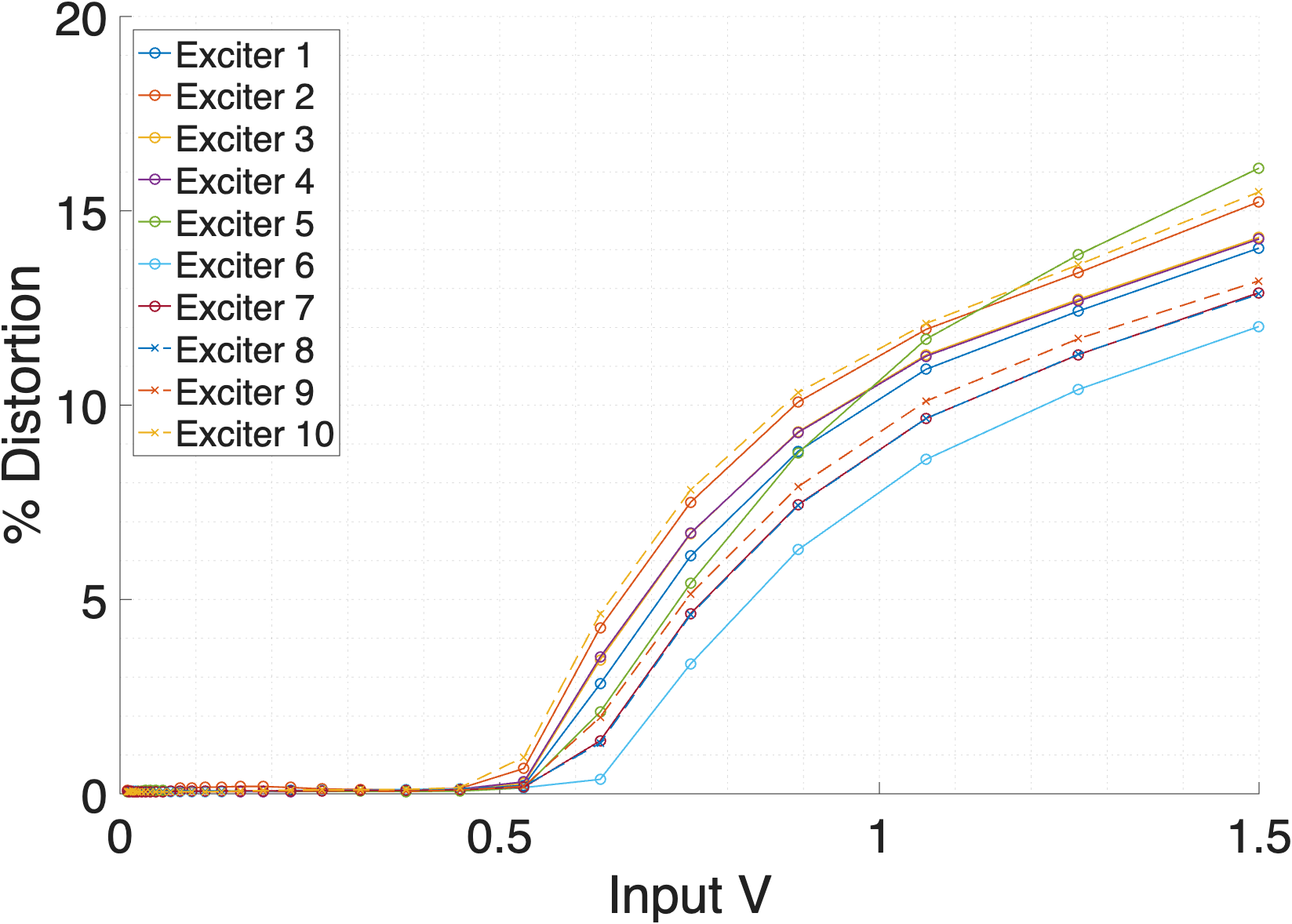}
\caption{THD of DAEX25CT-4}
\label{fig:DAEX25CT_4_THD}
\end{figure}

\begin{figure}[h]
\centering
\includegraphics[width= .9\linewidth]{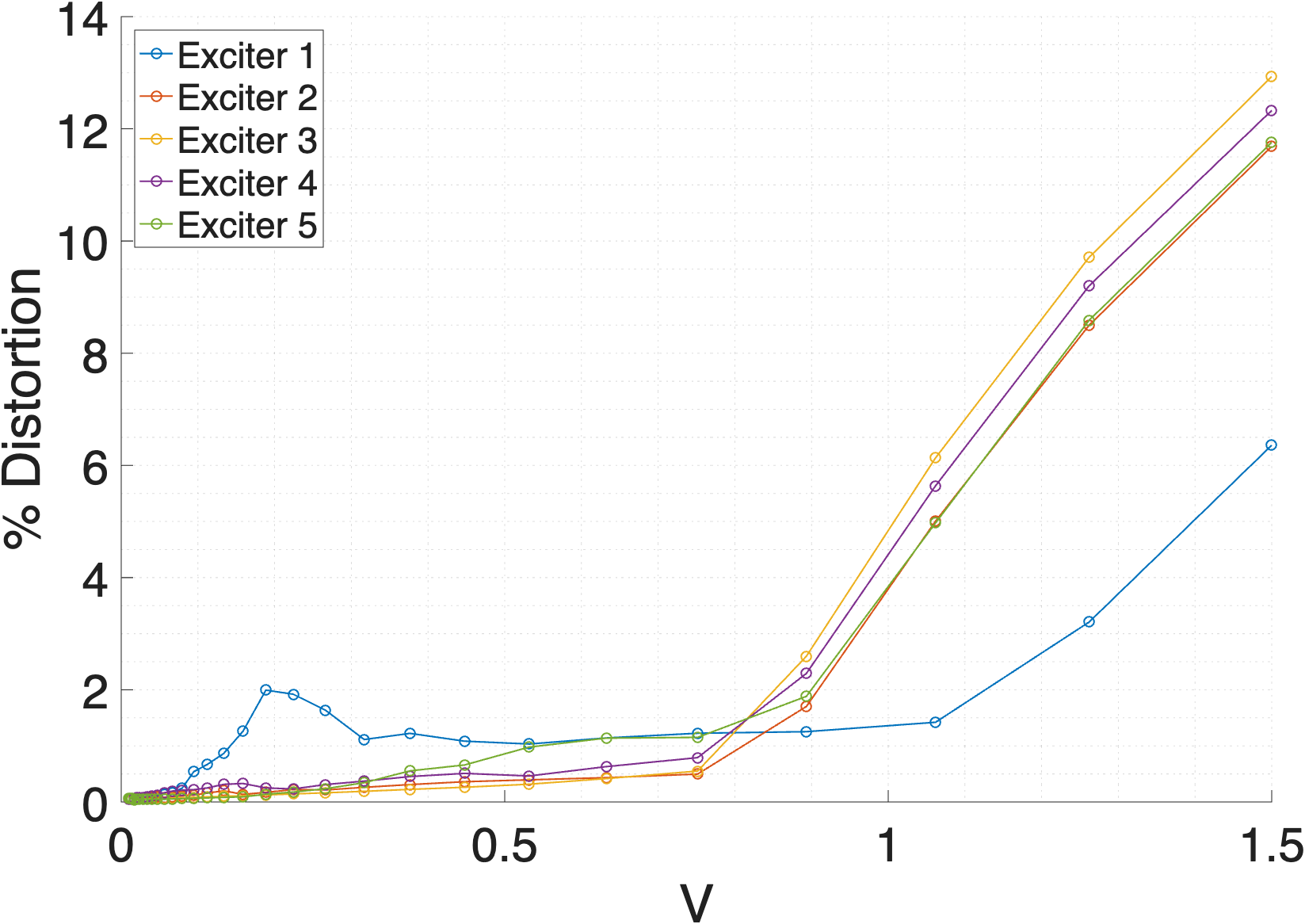}
\caption{THD of BS12K09}
\label{fig:BS12K09_THD}
\end{figure}

THD of the exciters can be seen in Figs.~\ref{fig:DAEX25FHE_4_THD}-\ref{fig:BS12K09_THD}. In each exciter model, THD rises with respect to input voltage, with single exciters rarely rising above 18$\%$ THD. This is well below the threshold of audibility for THD \cite{temme2012measurement}. However, in multi-exciter FPLs, THD may be aggregated to audible levels.

\subsection{Force v Displacement}
\begin{figure}[!h]
\includegraphics[width= .9\linewidth]{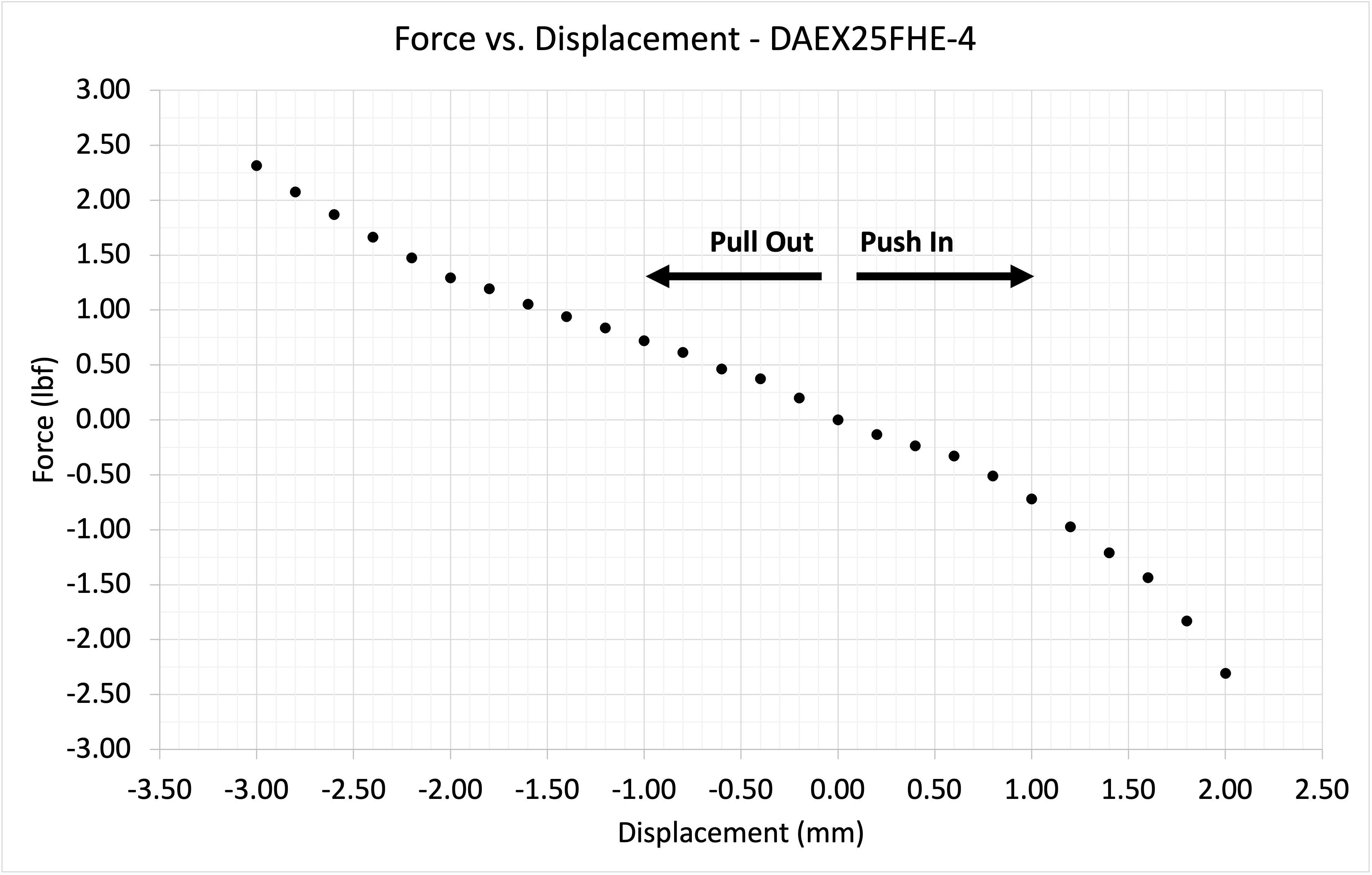}
\caption{Force (lbf) vs. Displacement (mm) of DAEX25FHE-4}
\label{fig:DAEX25FHE_4_FVD}
\end{figure}

\begin{figure}[h]
\centering
\includegraphics[width= .9\linewidth]{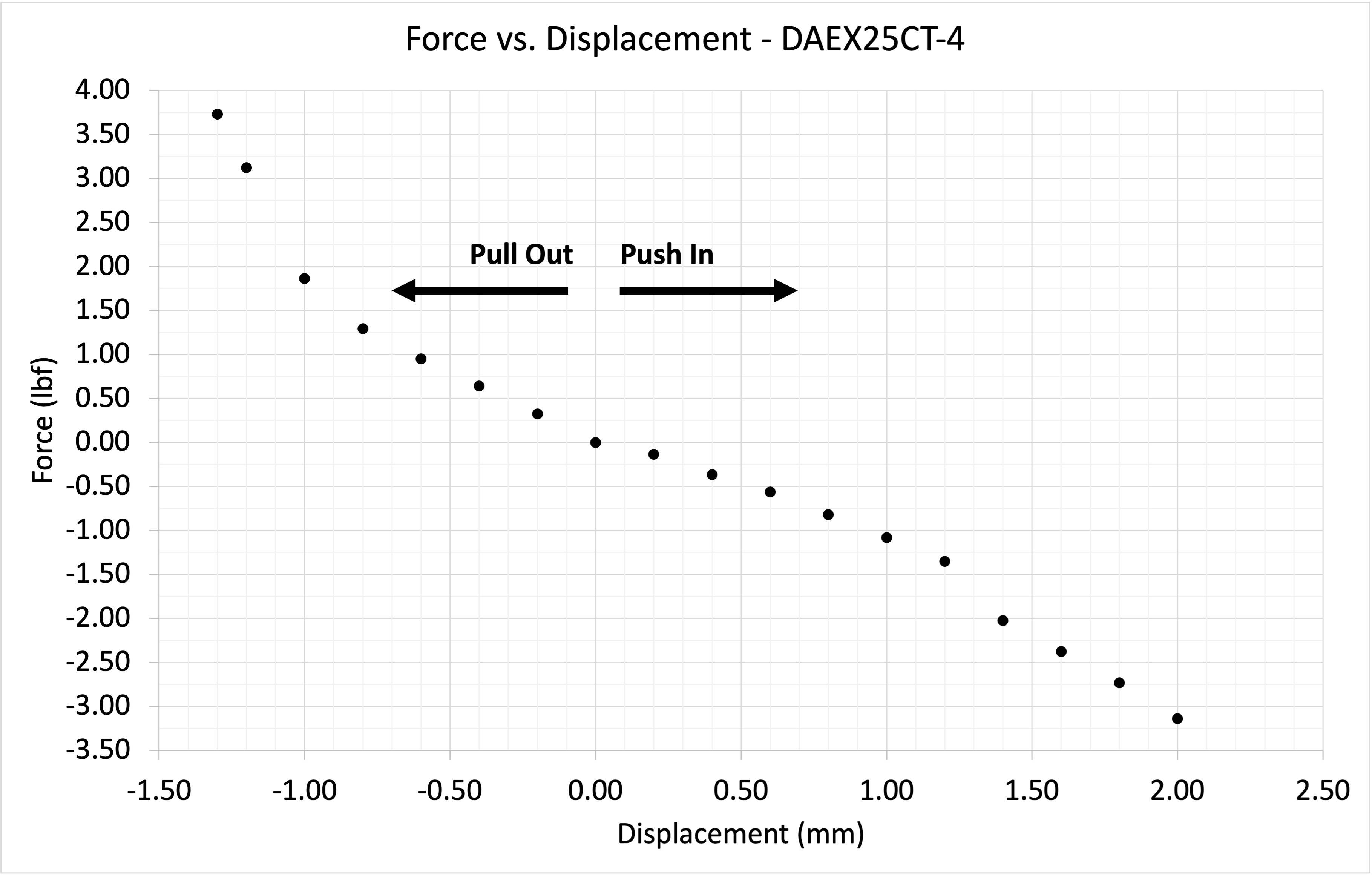}
\caption{Force (lbf) vs. Displacement (mm) of DAEX25CT-4}
\label{DAEX25CT_4_FVD}
\end{figure}

\begin{figure}[h]
\centering
\includegraphics[width= .9\linewidth]{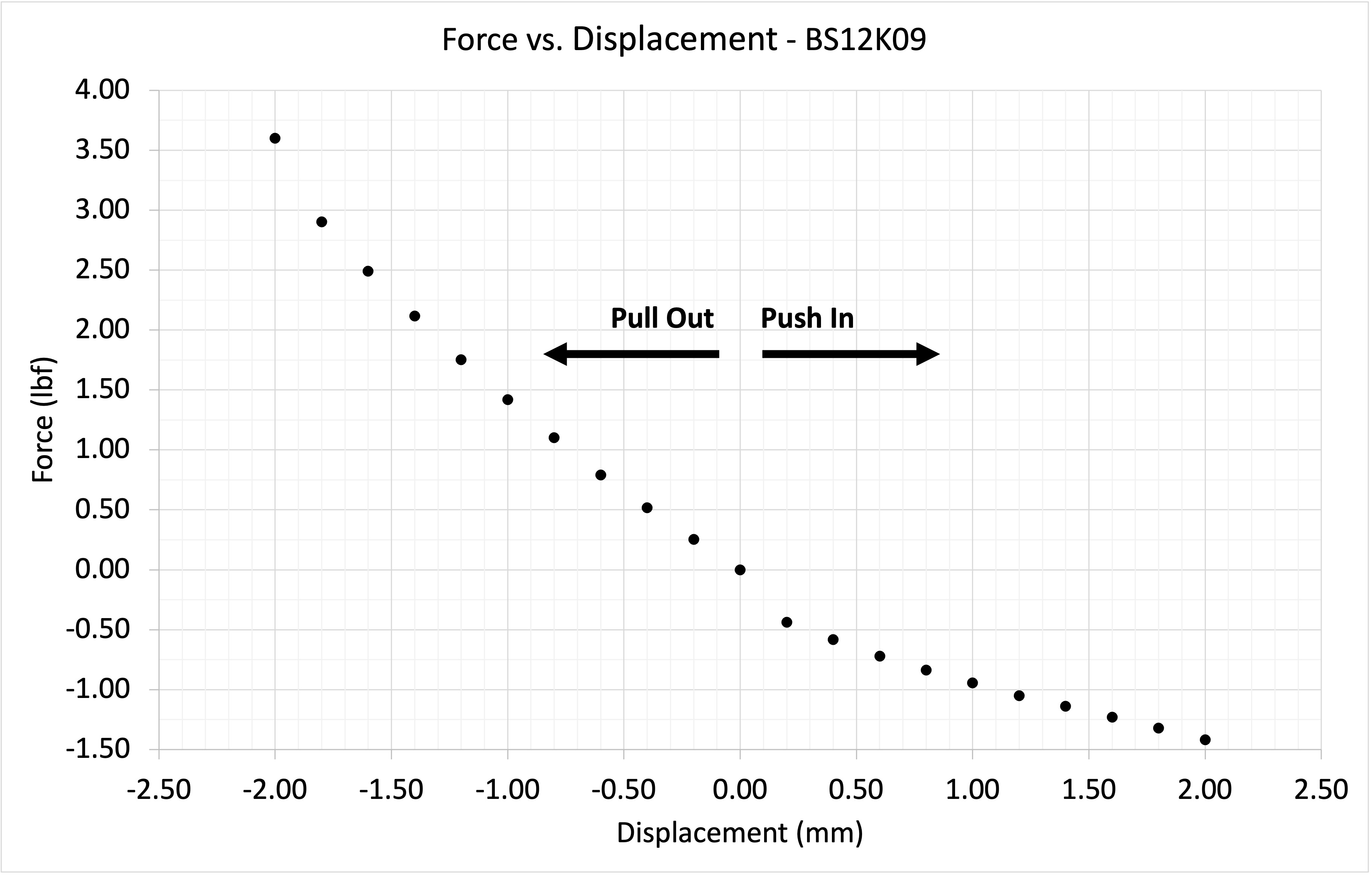}
\caption{Force (lbf) vs. Displacement (mm) of BS12K09}
\label{BS12K09_FVD}
\end{figure}

The force v. displacement of each exciter model is shown in Figs.~\ref{fig:DAEX25FHE_4_FVD}-\ref{BS12K09_FVD}. Since each graph is nonlinear, we can argue that the coil-gap is not equal length resulting in a misalignment. This is then reflected in the findings for THD, IMD, and Linearity.

\section{Exciter Matching}
\begin{figure}[h]
\centering
\includegraphics[width= .9\linewidth]{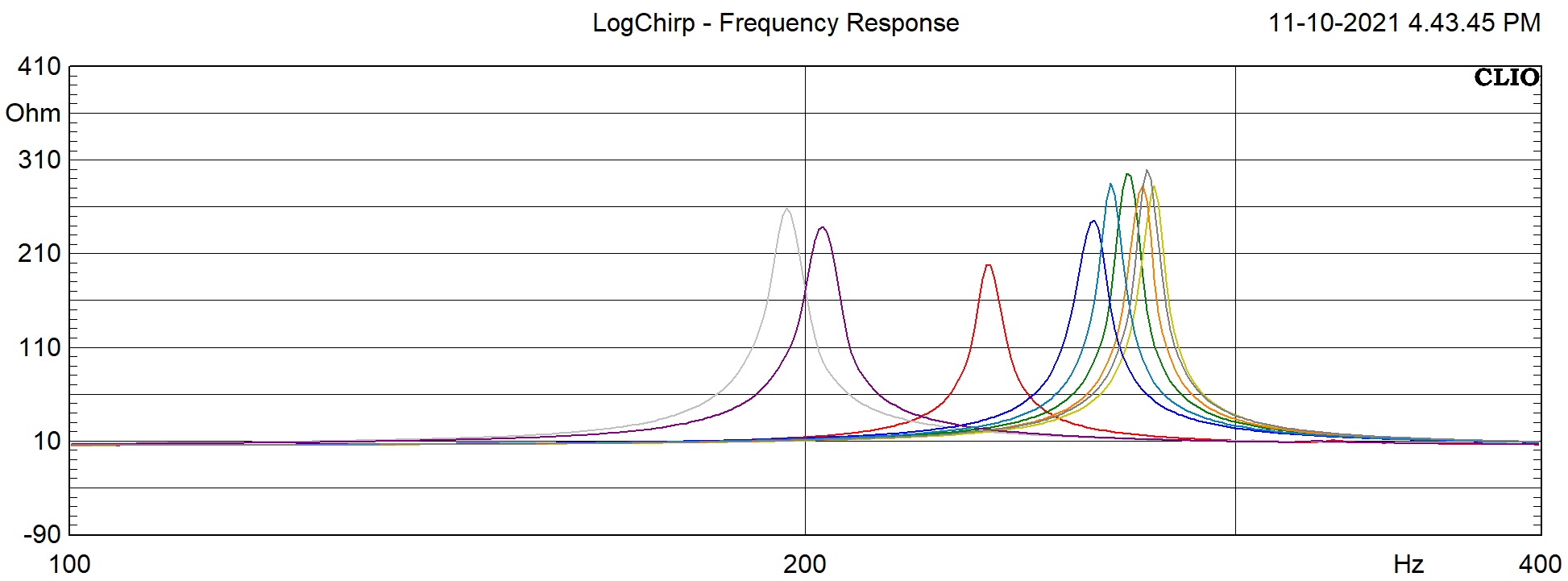}
\caption{The free-air impedance measurement of ten DAEX25FHE-4 exciters, where $f_0$ is shown to have nearly 100Hz of variability.}
\label{fig:Impedance_2}
\end{figure}

Generally, manufacturers of high-end loudspeakers do not match drivers, since they control the design and development of the loudspeaker from the ground up. However, they do perform tests on completed speakers to ensure they meet quality standards \cite{PMC}. Since the exciters measured above demonstrate high variability from unit to unit, multi-exciter FPL designs such as those described in \cite{heilemann2021evolution} should employ matched exciters to ensure consistent behavior. By using matched exciters, the FPL should exhibit less distortion in the far-field, as well as a smoother frequency response. 
\begin{figure}[!h]
\centering
\includegraphics[width= 0.5\linewidth]{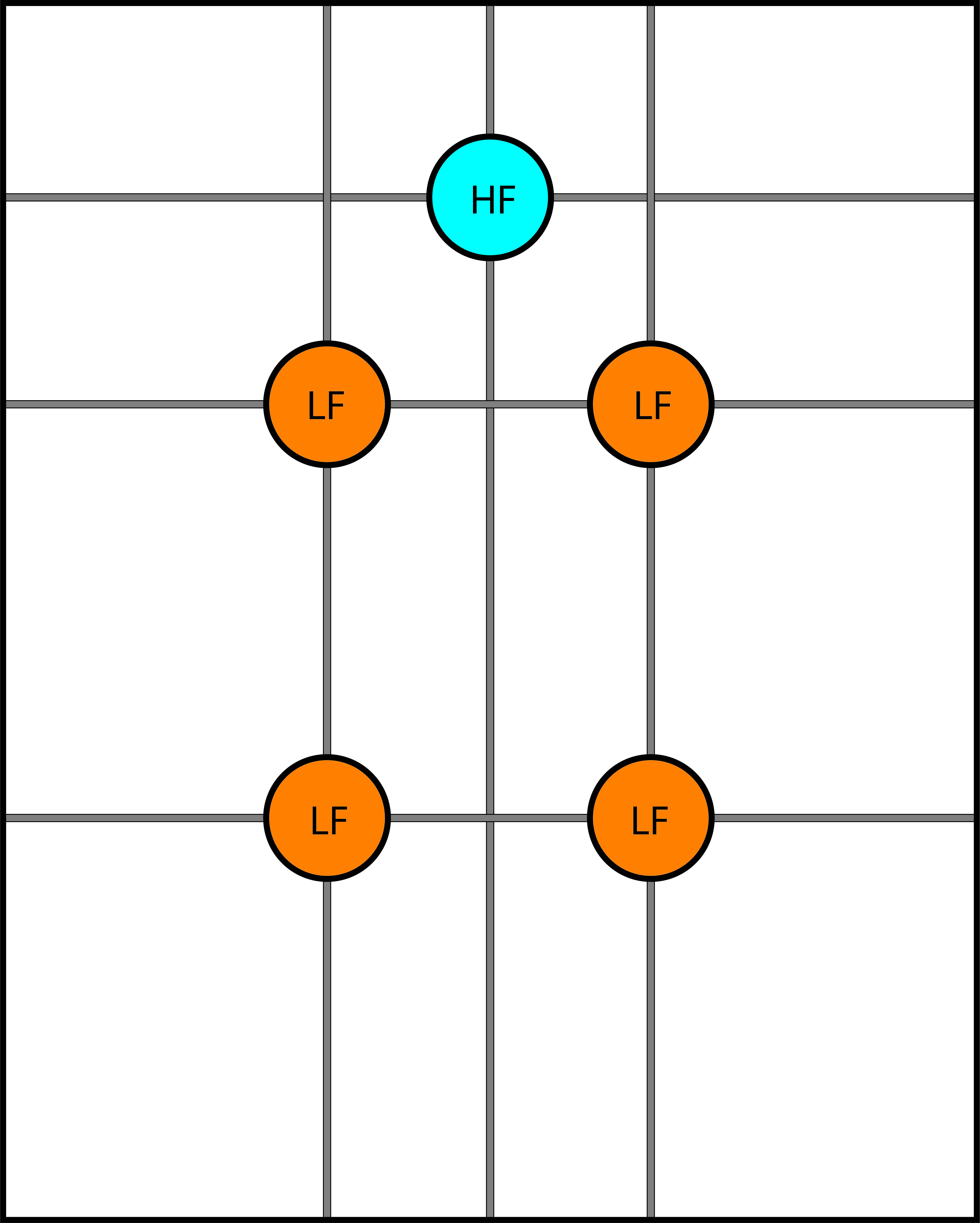}
\caption{Orientation of exciters on a 60cm x 120cm panel.}
\label{fig:DriverArray}
\end{figure}
Exciter matching was performed using free-air impedance measurements to group exciters with similar fundamental frequencies and response curves. Those with $f_0$ within a tolerance of $\pm$ 2Hz were matched together. An 60cm x 120cm FPL made of a damped PVC panel was measured by the CLIO 12 System in a Whisper Room using both matched and unmatched exciters. The drivers were arranged per recommendations in Anderson \textit{et al.} \cite{Anderson_Optimized}. The array can be seen in Fig.~\ref{fig:DriverArray}. DAEX25FHE-4 exciters were used, as they had the highest variability in response characteristics. The impedance of 10 exciters are shown in Fig.~\ref{fig:Impedance_2} from 100Hz~-~400Hz. Out of 10 exciters tested, 4 could be matched to each other within the tolerance range with resonances around 258Hz.

\begin{figure}[h]
\centering
\includegraphics[width= .9\linewidth]{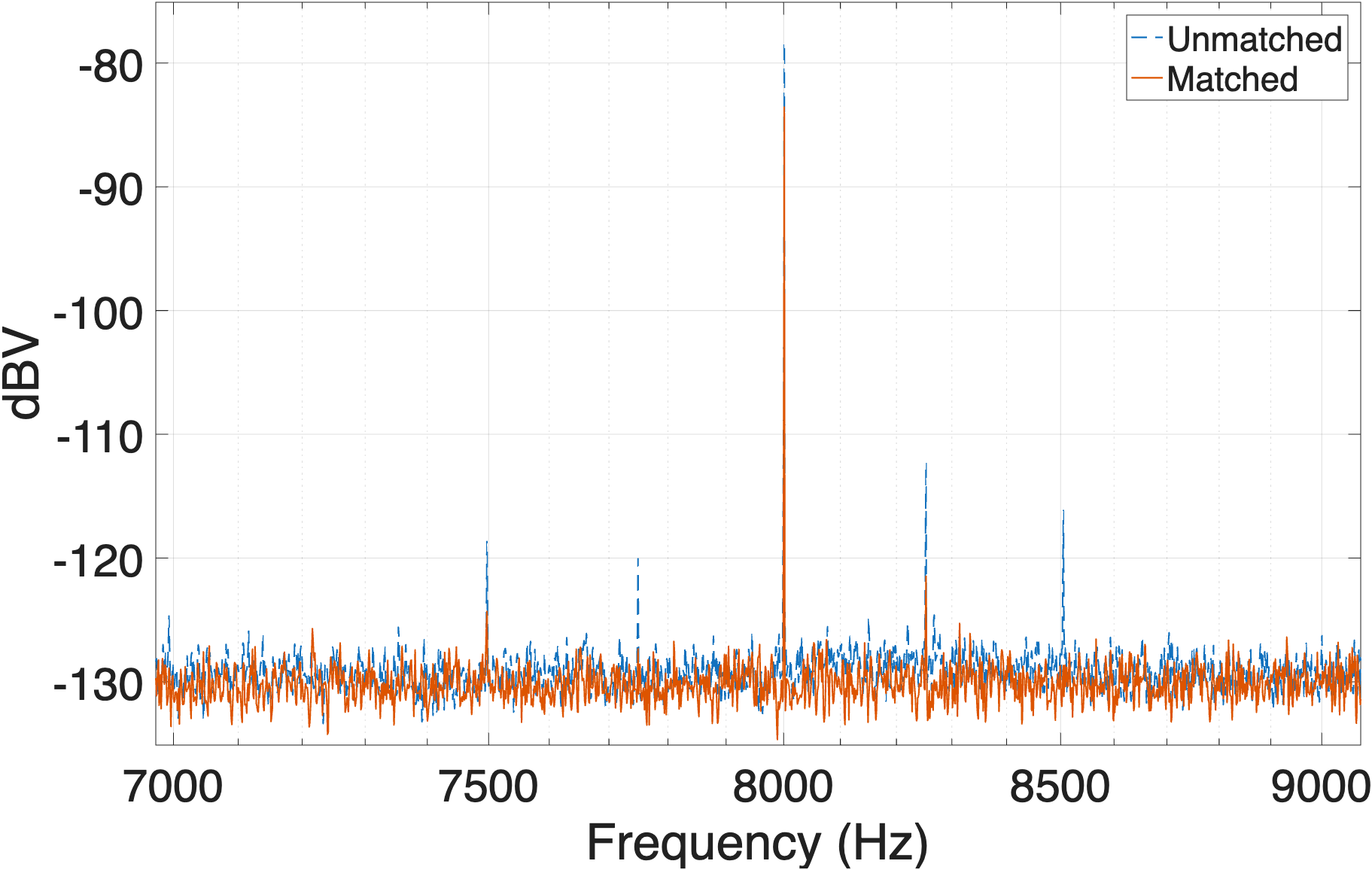}
\caption{Matched vs. Unmatched DIN IMD Measurement, centered around the 8kHz tone}
\label{fig:DINMatchUnmatchZoom}
\end{figure}

The consequences of matching exciters are especially evident in IMD measurements. The DIN measurement in Fig.~\ref{fig:DINMatchUnmatchZoom} shows reduced energy in the side-bands located at 7.5kHz, 7.75kHz, 8.25kHz, and 8.5kHz for the matched case compared to the unmatched case both in absolute gain reduction and gain reduction percentage relative to the 8kHz tone. 

\begin{figure}[h]
\centering
\includegraphics[width= .9\linewidth]{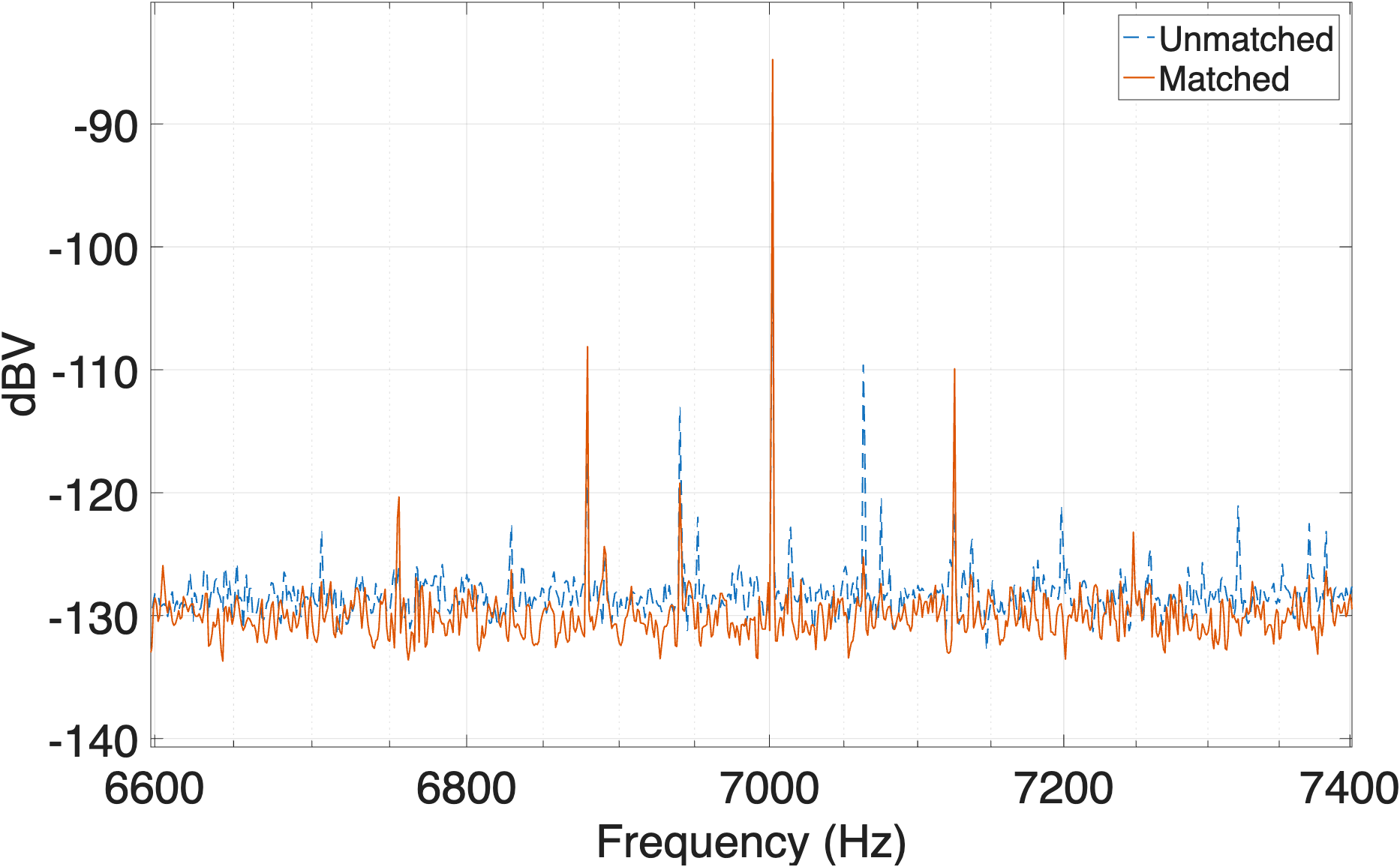}
\caption{Matched vs. Unmatched SMPTE IMD Measurement, centered around the 7kHz tone}
\label{fig:SMPTEMatchUnmatchZoom}
\end{figure}

IMD is also reduced with matched exciters for SMPTE measurements as shown in Fig.~\ref{fig:SMPTEMatchUnmatchZoom}. The expected side-bands at 6.94kHz and 7.06kHz are reduced with the matched drivers. However, it should be noted that the second harmonic side-bands at 6.88kHz and 7.12kHz show more energy in the matched case. This may be explained because the SMPTE measurement forces the exciters to excited near their mass-loaded resonant frequency. In this specific test case, an unmatched pair would actually score better because exciters with significantly different resonant frequencies would not emphasize harmonic distortion brought about when exciter is driven near resonance.

\section{Conclusion}

Commercially available exciters used for FPL design are shown to have high variability in performance metrics between units that affect loudspeaker performance. Two common issues were consistent with all exciter types. The first was  ringing in the frequency response. This may be caused by a weak surround on the exciter. Much like a traditional loudspeaker, if the surround stiffness is too weak, the cone will continue to push and pull after signal is removed, causing unwanted distortions throughout. Dampening these resonances with a stiffer surround, or perhaps a heavier cone could help improve exciter design and response.

The second was an anomaly in all exciters when a voltage is applied between 0.2V and 0.3V. The THD percentage increases stubstantially at this input voltage when the non-linear exciter behavior begins. One possibility for causing this issue may be that the voice coil gets stuck to or caught on the magnet at low voltages. Once enough voltage is applied, the voice coil is able to release.

A quality control check of linearity and/or force v. displacement to match exciters with comparable performance characteristics leads to improved loudspeaker performance. The methods for exciter testing in this work are not unlike those used by manufacturers of speakers and microphones who continually test their products for consistency and match. Given the widespread use of these exciters in FPL design, we hope that this work motivates improved quality control practices by both manufacturers and users, as reporting the make and model of exciter used in an experiment is not sufficient to accurately describe the response characteristics. 




\bibliographystyle{jaes}
\bibliography{refs}

@article{Flanagan1999,
   author = {Neil Harris and Sheila Flanagan},
   journal = {Audio Engineering Society Convention 106},
   title = {Loudness: A Study of the Subjective Difference between DML and Conventional Loudspeakers},
   year = {1999},
}

@article{klipsch1969modulation,
  title={Modulation Distortion in Loudspeakers},
  author={Klipsch, Paul W},
  journal={Journal of the Audio Engineering Society},
  volume={17},
  number={2},
  pages={194--206},
  year={1969},
  publisher={Audio Engineering Society}
}

@article{klipsch1970modulation,
  title={Modulation Distortion in Loudspeakers: Part 2},
  author={Klipsch, Paul W},
  journal={Journal of the Audio Engineering Society},
  volume={18},
  number={1},
  pages={29--33},
  year={1970},
  publisher={Audio Engineering Society}
}

@article{klipsch1972modulation,
  title={Modulation Distortion in Loudspeakers: Part 3},
  author={Klipsch, Paul W},
  journal={Journal of the Audio Engineering Society},
  volume={20},
  number={10},
  pages={827--828},
  year={1972},
  publisher={Audio Engineering Society}
}

@article{heilemann2021evolution,
  title={The Evolution and Design of Flat-Panel Loudspeakers for Audio Reproduction},
  author={Heilemann, Michael C and Anderson, David A and Roessner, Stephen and Bocko, Mark F},
  journal={Journal of the Audio Engineering Society},
  volume={69},
  number={1/2},
  pages={27--39},
  year={2021},
  publisher={Audio Engineering Society}
}

@inproceedings{Cecchi2019applying,
title={{Applying Sound Equalization to Vibrating Sound Transducers Mounted on Rigid Panels}},
author={Cecchi, Stefania and Terenzi, Alessandro and Piazza, Francesco and Bettarelli, Ferruccio},
year={2019},
month={oct},
booktitle={Journal of the Audio Engineering Society},
publisher={Paper 10233; AES Convention 147; October 2019},
number={10233},
organization={AES},
}

@article{10.1121/10.0036373,
    author = {Kleczkowski, Piotr and Makuch, Teresa and Król-Nowak, Aleksandra and Czesak, Karol},
    title = {Equalized distributed mode loudspeakers rival conventional ones in listening tests},
    journal = {The Journal of the Acoustical Society of America},
    volume = {157},
    number = {4},
    pages = {2528-2541},
    year = {2025},
    month = {04},
    doi = {10.1121/10.0036373},
    url = {https://doi.org/10.1121/10.0036373},
}

@article{anderson2017impulse,
author={Anderson, David and Heilemann, Michael and Bocko, Mark F.},
journal={Journal of the Audio Engineering Society},
title={Impulse and Radiation Field Measurements for Single Exciter Versus Exciter Array Flat-Panel Loudspeakers},
year={2017},
volume={},
number={},
pages={},
doi={},
month={october},
}

@inproceedings{Zenker2019rethinking,
title={{Rethinking Flat Panel Loudspeakers—An Objective Acoustic Comparison of Different Speaker Categories}},
author={Zenker, Benjamin and Merchel, Sebastian and Altinsoy, M. Ercan},
year={2019},
month={oct},
booktitle={Journal of the Audio Engineering Society},
publisher={Paper 10243; AES Convention 147; October 2019},
number={10243},
organization={AES},
}

@article{anderson2017flat-panel,
author={Anderson, David and Heilemann, Michael and Bocko, Mark F.},
journal={Journal of the Audio Engineering Society},
title={Flat-Panel Loudspeaker Simulation Model with Electromagnetic Inertial Exciters and Enclosures},
year={2017},
volume={65},
number={9},
pages={722-732},
doi={https://doi.org/10.17743/jaes.2017.0027},
month={september},
}

@article{beers1943frequency,
  title={Frequency-modulation distortion in loudspeakers},
  author={Beers, GL and Belar, H},
  journal={Proceedings of the IRE},
  volume={31},
  number={4},
  pages={132--138},
  year={1943},
  publisher={IEEE}
}

@article{MXN,
  Title                    = {Modal Crossover Networks for Flat-Panel Loudspeakers},
  Author                   = {David Anderson and Mark F. Bocko},
  Journal                  = {J. Audio Eng. Soc.},
  Year                     = {2016},
  Month                    = apr,
  Number                   = {4},
  Pages                    = {229-240},
  Volume                   = {64}
}

@article{klippel2001speaker,
  title={Speaker Auralization-Subjective Evaluation of Nonlinear Distortion},
  author={Klippel, Wolfgang},
  journal={PREPRINTS-AUDIO ENGINEERING SOCIETY},
  year={2001},
  publisher={Audio Engineering Society; 1999}
}

@article{thiele1971loudspeakers,
  title={Loudspeakers in vented boxes: Part 1},
  author={Thiele, Neville},
  journal={Journal of the Audio Engineering Society},
  volume={19},
  number={5},
  pages={382--392},
  year={1971},
  publisher={Audio Engineering Society}
}

@article{Anderson_Optimized,
  Title                    = {Optimized Driver Placement for Array-Driven Flat-Panel Loudspeakers},
  Author                   = {David Anderson and Michael C. Heilemann and Mark F. Bocko},
  Journal                  = {Arch. Acoust.},
  Year                     = {2017},
  Volume			= {42},
  Number			= {1},
  Pages			= {93-104}
}

@book{dickason2000loudspeaker,
  title={The Loudspeaker Design Cookbook},
  author={Dickason, Vance},
  year={2000},
  publisher={Audio Amateur Press Peterborough}
}

@conference{BankHarris,
  Title                    = {The Distributed Mode Loudspeaker-Theory and Practice},
  Author                   = {Bank, Graham and Harris, Neil},
  Booktitle                = {AES 13th UK Conference: Microphones \& Loudspeakers},
  Year                     = {1998},
  Type		    = {No. MAL-18},
  Month                    = mar
}

@conference{roessner2019evaluating,
		title = {Evaluating Listener Preference of Flat-Panel Loudspeakers},
		author = {Roessner, Stephen and Heilemann, Michael and Bocko, Mark F.},
		booktitle = {Audio Engineering Society Convention 147},
		month = {Oct},
		year = {2019},
		url = {http://www.aes.org/e-lib/browse.cfm?elib=20603}
		}

@conference{klippel2007optimal,
		title = {Optimal Design of Loudspeakers with Nonlinear Control},
		author = {Klippel, Wolfgang},
		booktitle = {Audio Engineering Society Conference: 32nd International Conference: DSP For Loudspeakers},
		month = {Sep},
		year = {2007},
		url = {http://www.aes.org/e-lib/browse.cfm?elib=14214}
		}

@inproceedings{olive2004multiple2,
  title={A Multiple Regression Model for Predicting Loudspeaker Preference Using Objective Measurements: Part II-Development of the Model},
  author={Olive, Sean E},
  booktitle={Audio Engineering Society Convention 117},
  year={2004},
  organization={Audio Engineering Society}
}

@inproceedings{olive2004multiple1,
  title={A Multiple Regression Model for Predicting Loudspeaker Preference Using Objective Measurements: Part I-Listening Test Results},
  author={Olive, Sean E},
  booktitle={Audio Engineering Society Convention 116},
  year={2004},
  organization={Audio Engineering Society}
}

@inproceedings{klippel2005loudspeaker,
  title={Loudspeaker nonlinearities--causes, parameters, symptoms},
  author={Klippel, Wolfgang},
  booktitle={Audio Engineering Society Convention 119},
  year={2005},
  organization={Audio Engineering Society}
}

@inproceedings{HeilemannListenerPreference,
  title={Quantifying Listener Preference of Flat-Panel Loudspeakers},
  author={Heilemann, Michael and Anderson, David and Roessner, Stephen and Bocko, Mark F},
  booktitle={Audio Engineering Society Convention 145},
  year={2018},
  organization={Audio Engineering Society}
}

@inproceedings{temme2012measurement,
  title={Measurement of harmonic distortion audibility using a simplified psychoacoustic model},
  author={Temme, Steve and Brunet, Pascal and Qarabaqi, Parastoo},
  booktitle={Audio Engineering Society Convention 133},
  year={2012},
  organization={Audio Engineering Society}
}

@misc{metzler1993,
  title={Measurement of Non-Linearity, Audio Precision Audio Measurement Handbook, Audio Precision},
  author={Metzler, Bob},
  year={1993},
  publisher={Inc}
}

@misc{lane1930sound,
  title={Sound radiator},
  author={Lane, Clarence E},
  year={1930},
  month=aug,
  publisher={Google Patents},
  note={{US} Patent 1,773,910}
}

@misc{FSAMSL,
  author = {FSA-MSL},
  title  = {FSA-MSL Force / Displacement Tester},
  year   = {2021},
  url    = {https://www.checkline.com/product/FSA-MSL}
}

@misc{PMC,
  author       = "{PMC Speakers}",
  title        = "{Research and Development}",
  year         = {2022},
  howpublished = {\url{https://pmc-speakers.com/technology/research-and-development/}},
  note         = {Accessed: 2022-09-16}
}

@misc{Hiromi,
  title={Loudspeaker},
  author={Hiromi, Sotome},
  year={1971},
  month= may,
  day = {11},
  note={{US} Patent 3,578,104 }
}

@misc{AzimaPatent,
  title={Acoustic device},
  author={Azima, H.},
  year={2004},
  month=feb # "~17",
  note={{US} Patent 6,694,038}
}

@phdthesis{HarrisThesis,
  title={The acoustics and psychoacoustics of the distributed-mode loudspeaker (DML).},
  author={Harris, Neil John},
  year={2001},
  school={University of Essex}
}

@techreport{DIN45403-1,
author = {DIN},
title = {Measurement of non-linear distortions in electoacoustics},
pages = {2},
year = {1963},
institution = {Deutsches Institut Normung},
howpublished = {https://www.dinmedia.de/en/standard/din-45403-1/1520377}
}

@techreport{SMPTE-RP120-2005,
author = {SMPTE},
title = {Measurement of Intermodulation Distortion in Analog Audio Systems},
pages = {2},
year = {2005},
institution = {Society of Motion Picture and Television Engineers},
howpublished = {https://pub.smpte.org/latest/rp120/rp0120-2005_stable2015.pdf}
}

\end{document}